\documentclass[11pt]{emulateapj}
\usepackage{natbib}
\usepackage{graphicx}
\usepackage{amsmath}
\usepackage{amssymb}
\def\jgr{J. Geophys. Res. }

\def\aj{Astron. J. }

\def\chf{CH$_4$ }
\def\chfx{CH$_4$}
\def\chisq{$\chi^2$ }

\def\icm{cm$^{-1}$ }
\def\icmx{cm$^{-1}$}

\def\Wm2{W/m$^2$}
\def\Wpm2sr{Wm$^{-2}sr^{-1}$}
\def\deg{$^\circ$ }
\def\degx{$^\circ$}

\def\mum{$\mu$m }
\def\mumx{$\mu$m}

\def\nhfsh{NH$_4$SH }

\begin{document}
\title{Methane depletion in both polar regions
of Uranus inferred from HST/STIS\footnotemark[\dag] and Keck/NIRC2 observations}
\author{L.~A. Sromovsky\altaffilmark{1}, E. Karkoschka\altaffilmark{2}, P.~M. Fry\altaffilmark{1},
H.\ B. Hammel\altaffilmark{3}$^,$\altaffilmark{4}, I. de Pater\altaffilmark{5}, and K. Rages\altaffilmark{6}}
\altaffiltext{1}{University of Wisconsin - Madison, Madison WI 53706}
\altaffiltext{2}{University of Arizona, Tucson AZ 85721, USA}
\altaffiltext{3}{AURA, 1212 New York Ave. NW, Suite 450, Washington, DC 20005, USA}
\altaffiltext{4}{Space Science Institute, Boulder, CO 80303, USA}
\altaffiltext{5}{University of California, Berkeley, CA 94720, USA}
\altaffiltext{6}{SETI Institute, Mountain View, CA 94043, USA}
\altaffiltext{\dag}{Based in part on observations with the NASA/ESA Hubble Space
Telescope obtained at the Space Telescope Science Institute, which is
operated by the Association of Universities for Research in Astronomy,
Incorporated under NASA Contract NAS5-26555.} 

\slugcomment{Journal reference: Icarus 238 (2014) 137-155.}
\begin{abstract}

From Space Telescope Imaging Spectrograph (STIS) observations of
Uranus in 2012, when good views of its north polar regions were
available, we found that the methane volume mixing ratio declined from
about 4\% at low latitudes to about 2\% at 60\deg N and beyond. This
depletion in the north polar region of Uranus in 2012 is similar in
magnitude and depth to that found in the south polar regions in 2002.
This similarity is remarkable because of the strikingly different
appearance of clouds in the two polar regions: we have never seen any
obvious signs of convective activity in the south polar region, while
the north has been peppered with numerous small clouds thought to be
of convective origin.  Keck and Hubble Space Telescope imaging
observations close to equinox at wavelengths of 1080 nm and 1290 nm,
with different sensitivities to methane and hydrogen absorption but
similar vertical contribution functions, imply that the depletions
were simultaneously present in 2007, and at least their gross
character is probably a persistent feature of the Uranus
atmosphere. The depletion appears to be mainly restricted to the upper
troposphere, with the depth increasing poleward from about 30\deg N,
reaching $\sim$4 bars at 45\deg N and perhaps much deeper at 70\deg N,
where it is not well constrained by our observations. The latitudinal
variations in degree and depth of the depletions are important
constraints on models of meridional circulation.  Our observations are
qualitatively consistent with previously suggested circulation cells
in which rising methane-rich gas at low latitudes is dried out by
condensation and sedimentation of methane ice particles as the gas
ascends to altitudes above the methane condensation level, then is
transported to high latitudes, where it descends and brings down
methane depleted gas.  Since this cell would seem to inhibit formation
of condensation clouds in regions where clouds are actually inferred
from spectral modeling, it suggests that sparse localized convective
events may be important in cloud formation. A more complex meridional
circulation pattern may be necessary to reproduce the observed cloud
distribution,
but microwave observations appear to be most compatible with a single
deep circulation cell.  The small-scale latitudinal variations we
found in the effective methane mixing ratio between 55\deg N and
82\deg N have significant inverse correlations with zonal mean
latitudinal variations in cloud reflectivity in near-IR Keck images
taken before and after the HST observations. If the \chf/H$_2$
absorption ratio variations are interpreted as local variations in
para fraction instead of methane mixing ratio, we find that
downwelling correlates with reduced cloud reflectivity.  While there has
been no significant secular change in the brightness of Uranus at
continuum wavelengths between 2002 and 2012, there have been
significant changes at wavelengths sensing methane and/or hydrogen
absorption, with the southern hemisphere darkening considerably
between 2002 and 2012, by $\sim$25\% at mid latitudes near 827 nm, for
example, while the northern hemisphere has brightened by $\sim$25\% at
mid latitudes at the same wavelength.

\end{abstract}
\keywords{Uranus, Uranus Atmosphere; Atmospheres, composition; Atmospheres, dynamics}

\maketitle
\shortauthors{Sromovsky et al.} 
\shorttitle{Polar methane depletion on Uranus.}


\section{Introduction}

\begin{figure*}\centering
\includegraphics[width=5.2in]{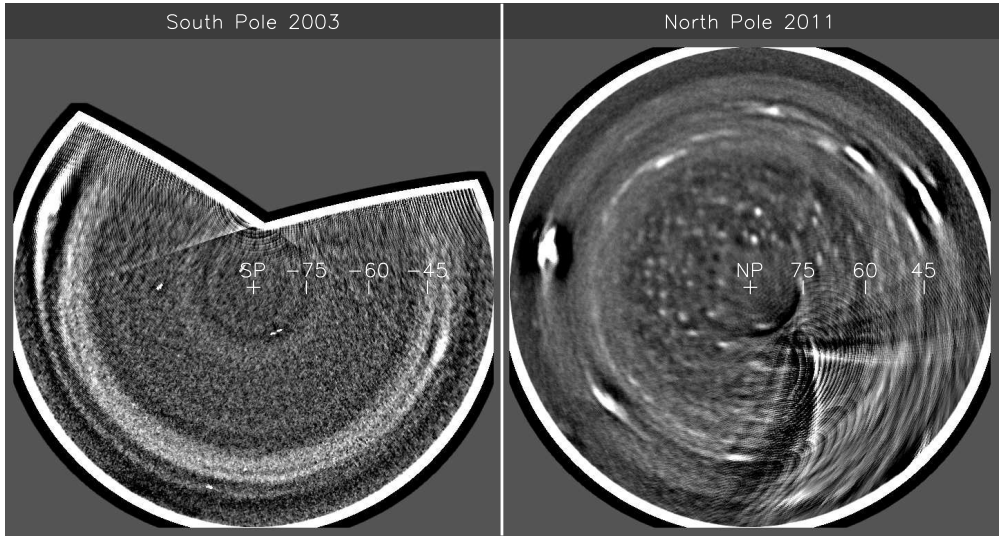}
\caption{Shift-and-add high-pass filtered projections of Uranus' 2003
  South (L) and 2011 North (R) polar regions reveal a large asymmetry
  in cloud features poleward of 45\degx, suggesting very different
  degrees of polar convection, a likely seasonal effect
  \citep{Sro2012polar}. The images were obtained with the Keck 2
  telescope and NIRC2 instrument using an H filter (1.62 \mum central
  wavelength). The small bright features in the polar region in the
  2011 image are 1-2\% brighter than their surroundings. The features
  in the lower right quadrant of that image are heavily distorted by
  projection artificats.}
\label{Fig:polar}
\end{figure*}

Uranus experiences the solar system's largest fractional seasonal
forcing because its spin axis has a 98\deg inclination to its orbital
plane. It is thus not surprising to see a time-dependent north-south
asymmetry in Uranus' cloud structure.  An especially interesting
asymmetry noted by \cite{Sro2012polar} is the continued complete
absence of discrete cloud features south of 45\deg S, while numerous
discrete cloud features have been observed north of 45\deg N in recent
near-IR H-filter Keck images (Fig.\ \ref{Fig:polar}). Voyager imaging
in 1986 recorded no bright cloud features between the south pole and
45\deg S.  Nor were any seen in near-IR Hubble Space Telescope (HST)
images taken with the Near-IR Camera and Multi-Object Spectrometer
(NICMOS) beginning in 1997, nor in Keck images beginning in 2003.
Some mechanism appears to be inhibiting convection at high southern
latitudes that is not present at high northern latitudes. A very
significant and possibly related result came from analysis of 2002
Space Telescope Imaging Spectrometer (STIS) observations by
\cite{Kark2009IcarusSTIS}, subsequently referenced as {\bf KT2009},
and confirmed by the analysis of \cite{Sro2011occult}. They found a
strong depletion of methane in the upper troposphere (down to a few
bars at least) at high southern latitudes, suggesting a downwelling
flow at these latitudes, which would tend to inhibit convective cloud
formation. This raised the possibility of a connection between methane
depletion and a lack of discrete cloud features, suggesting that high
northern latitudes, where discrete clouds are seen, might not be
depleted in methane. If so, the methane depletion might be a seasonal
effect. In 2002, five years before equinox, the sub-observer latitude
was 21.4\deg S and the north polar region was not visible, so that
testing this hypothesis would require new observations when the north
polar region was exposed to view.  We thus proposed new STIS
observations of Uranus, which were obtained in late September 2012
(HST program 12894, Sromovsky, PI), five years after equinox, when the
sub-observer latitude was 19.5\deg N. The analysis of these new
observations is the primary topic in what follows.

\begin{figure}\centering
\includegraphics[width=2.8in]{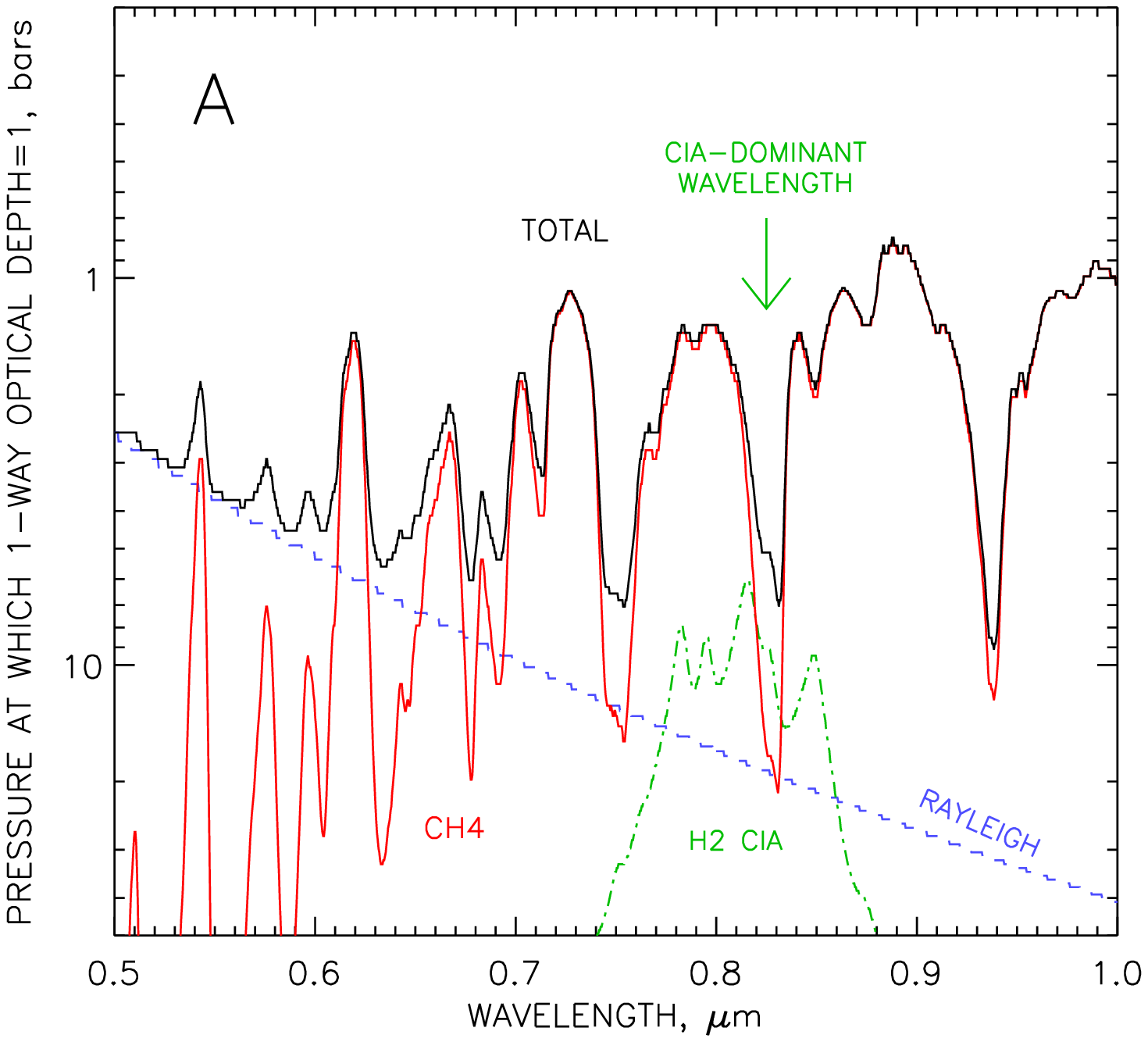}
\includegraphics[width=3.4in]{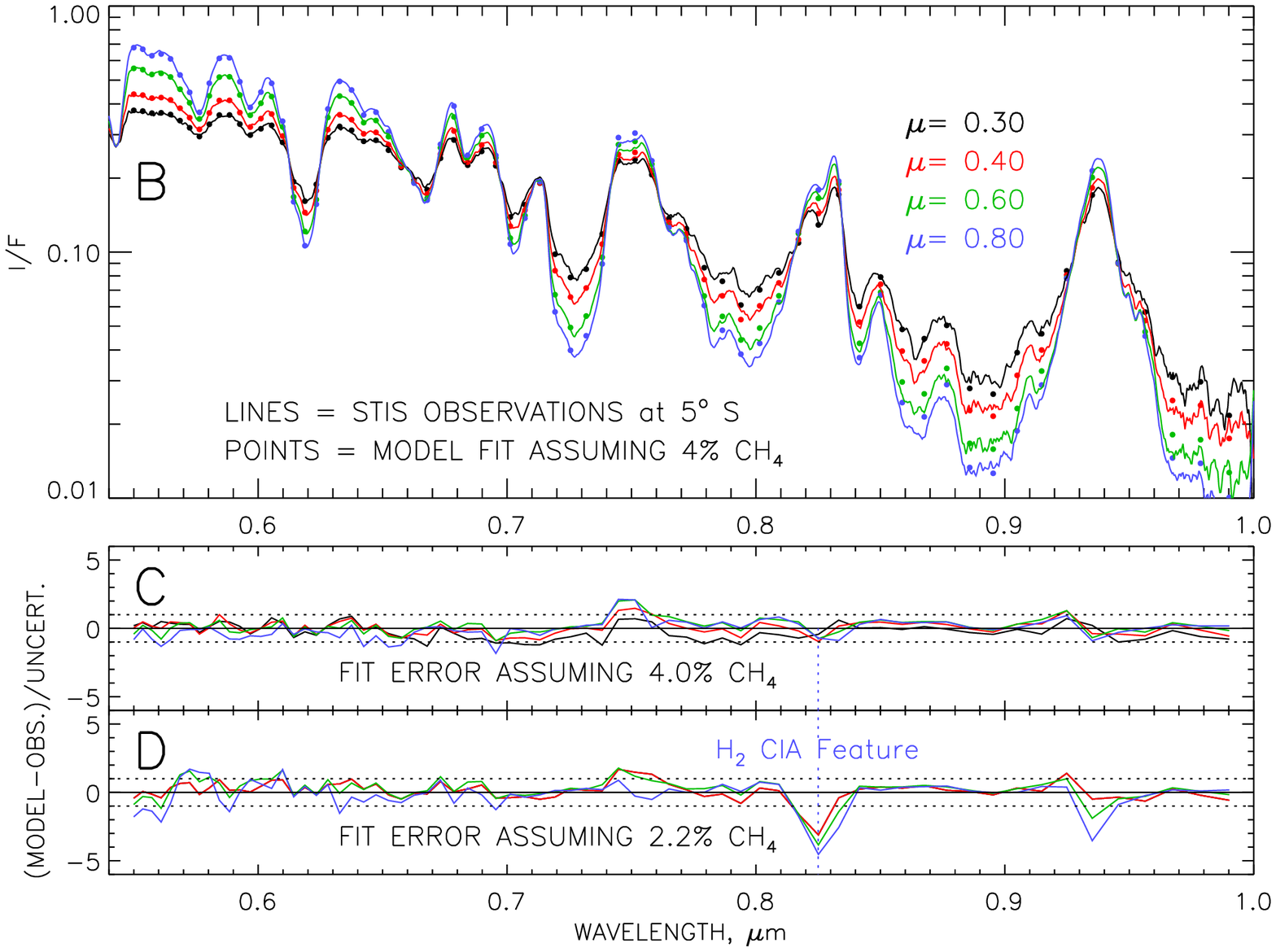}
\caption{A: Penetration depth vs. wavelength as limited by different
  opacity sources assuming the F1 methane profile and absorption
  models discussed in Sec.\ \ref{Sec:radtran}. Note the greater
  importance of H$_2$ CIA near 825 nm. The dark solid curve includes
  all absorbers, while the other curves are for single absorbers in
  isolation. B: Example model fit for the F1 profile of \cite{Sro2011occult} with a deep
  methane volume mixing ratio of 4\%, and fit errors for deep methane
  mixing ratios of 4\% (C) and 2.2\% (D).  The fit errors near 825 nm
  are a sensitive indicator of the mixing ratio of \chf to H$_2$.}
\label{Fig:ccdpendepth}
\end{figure}

Constraining the mixing ratio of \chf on Uranus is based on
differences in the spectral absorption of \chf and H$_2$, illustrated
by the penetration depth plot of Fig.\ \ref{Fig:ccdpendepth}A.
Methane absorption dominates at most wavelengths, but hydrogen's
Collision Induced Absorption (CIA) is relatively more important in a
narrow spectral range near 825 nm.  Model calculations that don't
have the correct ratio of methane to hydrogen lead to a relative
reflectivity mismatch near this wavelength.  An example is shown in
Fig.\ \ref{Fig:ccdpendepth}B-D, in which model calculations are
compared to 2002 STIS observations assuming methane profiles with
2.2\% and 4.0\% deep volume mixing ratios.  The two assumptions lead
to very different errors in the vicinity of 825 nm, clearly indicating
that the larger mixing ratio is a better choice.
\cite{Kark2009IcarusSTIS} used this spectral constraint to infer a
methane mixing ratio of 3.2\% at low latitudes, but dropping to 1.4\%
at high southern latitudes.
\cite{Sro2011occult} analyzed the same data set, but used only
temperature and mixing ratio profiles that were consistent with the
\cite{Lindal1987} refractivity profiles.  They confirmed the depletion
but inferred a somewhat higher mixing ratio of 4\% at low latitudes
and found that better fits were obtained if the high latitude
(down to $\sim$2-4 bars).  Subsequently, 2009 groundbased spectral
observations at the NASA Infrared Telescope Facility (IRTF) using the SpeX
instrument, which provided coverage of the key 825-nm
spectral region, were used by \cite{Tice2013} to infer that both polar
regions were weakly depleted in methane, but they inferred lower
methane mixing ratios, smaller latitudinal variations, and higher uncertainties than the
STIS-based analysis of KT2009 and \cite{Sro2011occult}. These lower IRTF-based
values might be a result of lower spatial resolution combined with worse
view angles into the polar regions than obtained by HST observations.

In the following we begin with a description of our 2012 HST/STIS
observations, and the complex data reduction and calibration
procedures.  We then describe the relatively direct implications of
latitudinal spectral variations and simplified model results. Finally,
we describe our radiation transfer calculation methods and the results
of constraining the cloud structure and methane distributions to fit
the observed spectra.  We show that the methane depletion is indeed
present in both polar regions but of significantly different
character.  The results provide no obvious explanation for the
asymmetry in polar cloud structure, but do raise some important
questions about cloud formation. The important implications for
meridional circulation models are discussed before summarizing our
results.

\section{Observations}

Our 2012 observations used four HST orbits, three of them  
devoted to STIS spatial mosaics and one orbit to Wide Field Camera 3 (WFC3) support
imaging. The STIS observations were taken on 27-28 September 2012 and
the WFC3 observations on 30 September 2012. Observing conditions and
exposures are summarized in Table\ \ref{Tbl:sci_obs}.

\begin{table*} \centering
\caption{Science exposures from HST program 12894. All STIS spectra
  used the 52$''$$\times$0.1$''$ slit.\label{Tbl:sci_obs}}
\begin{tabular}{|c | c | c | c | c | r | r | r|}
\hline
  &Start      & Start      & Instrument & Filter or & Exposure        & No. of & Phase\\[0.in]
Orbit &Date (UTH) & Time (UTH) &            & Grating   & (sec)  & Exp.      & Angle (\degx)\\
\hline
1 &2012-09-27 & 21:22:29   & STIS       & MIRVIS    & 10.1      & 1  &  0.09  \\[0.in]
1 &2012-09-27 & 21:38:11   & STIS       & G430L     & 70.0      & 13 &  0.09  \\[0.in]
2 &2012-09-27 & 22:56:43   & STIS       & G750L     & 86.0      & 18 &  0.08  \\[0.in]
3 &2012-09-28 & 00:32:26   & STIS       & G750L     & 86.0      & 18 &  0.08  \\[0.in]
4 &2012-09-30 & 22:44:50   & WFC3       & F336W     & 30.0      & 1  &  0.09  \\[0.in]
4 &2012-09-30 & 22:46:35   & WFC3       & F467M     & 16.0      & 1  &  0.09  \\[0.in]
4 &2012-09-30 & 22:48:15   & WFC3       & F547M     & 6.0       & 1  &  0.09  \\[0.in]
4 &2012-09-30 & 22:49:39   & WFC3       & F631N     & 65.0      & 1  &  0.09  \\[0.in]
4 &2012-09-30 & 22:52:08   & WFC3       & F665N     & 52.0      & 1  &  0.09  \\[0.in]
4 &2012-09-30 & 22:54:15   & WFC3       & F763M     & 26.0      & 1  &  0.09  \\[0.in]
4 &2012-09-30 & 22:56:02   & WFC3       & F845M     & 35.0      & 1  &  0.09  \\[0.in]
4 &2012-09-30 & 22:57:56   & WFC3       & F953N     & 250.0     & 1  &  0.09  \\[0.in]
4 &2012-09-30 & 23:04:27   & WFC3       & FQ889N    & 450.0     & 1  &  0.09  \\[0.in]
4 &2012-09-30 & 23:16:02   & WFC3       & FQ937N    & 160.0     & 1  &  0.09  \\[0.in]
4 &2012-09-30 & 23:23:05   & WFC3       & FQ727N    & 240.0     & 1  &  0.09  \\
\hline
\end{tabular}
\vspace{0.1in}
\parbox{6.in}{On September 28 the sub-observer planetographic latitude
  was 19.5\deg S, the observer range was 19.0613 AU
  (2.851530$\times$10$^9$ km), and the equatorial angular diameter of
  Uranus was 3.6976 arcseconds.}
\end{table*}

\vspace{-0.1in}
\subsection{STIS spatial mosaics.} 
\vspace{-0.05in}
STIS observations used the G430L and G750L gratings and the CCD
detector, which has $\sim$0.05 arcsecond square pixels covering a
nominal 52$''\times$52$''$ square field of view (FOV) and a
spectral range from $\sim$200 to 1030 nm \citep{Hernandez2012}.
Using the 52$''\times$0.1$''$ slit, the resolving power varies from 500
to 1000 over each wavelength range due to fixed wavelength dispersion
of the gratings. Observations had to be carried out within a few days of Uranus
opposition (29 September 2012) when the telescope roll angle could be
set 
to orient the STIS slit parallel
to the spin axis of Uranus. 

One STIS orbit produced a mosaic of half of Uranus using the CCD
detector, the G430L grating, and 52$''\times$0.1$''$ slit. The G430L
grating covers 290 to 570 nm with a 0.273 nm/pixel dispersion. The
slit was aligned with Uranus' rotational axis, and stepped from the
evening limb to the central meridian in 0.152 arcsecond increments
(because the planet has no high spatial resolution center-to-limb
features at these wavelengths we used interpolation to fill in missing
columns of the mosaic). Two additional STIS orbits were used to mosaic
the planet with the G750L grating and 52$''\times$0.1$''$ slit
(524-1027 nm coverage with 0.492 nm/pixel dispersion), with limb to
central meridian stepping at 0.0569 arcsecond intervals. This was the
same procedure that was used successfully for HST program 9035 in 2002
(E. Karkoschka, P.I.).  As Uranus' equatorial radius was 1.85
arcseconds when observations were performed, stepping from one step
off the limb to the central meridian required 13 positions for the
G430L grating (at 0.152 arcseconds/step) and 36 for the G750L grating
(at 0.0569 arcseconds/step). Two orbits were needed to complete the
G750L grating observations, spanning a total time of 2 h 17 m, during which Uranus rotated
47\degx. This rotation was not a problem because of the high degree of
zonal symmetry of Uranus and because our analysis rejected any small
scale deviations from it, such as rare discrete cloud features.

Exposure times were similar to those used in the 2002 program, with
70-second exposures for G430L and 86-second exposures for G750L
gratings, using the 1 electron/DN gain setting. These exposures
yielded single-pixel signal-to-noise ratios of around 10:1 at 300 nm,
$>$ 40:1 from around 400 to 700 nm, and decreasing to around 20:1
(methane windows) to $<$ 10:1 (methane absorption bands) at 1000 nm.

\subsection{Supporting WFC3 imaging.}\label{Sec:synth}

\begin{figure*}\centering
\includegraphics[width=6in]{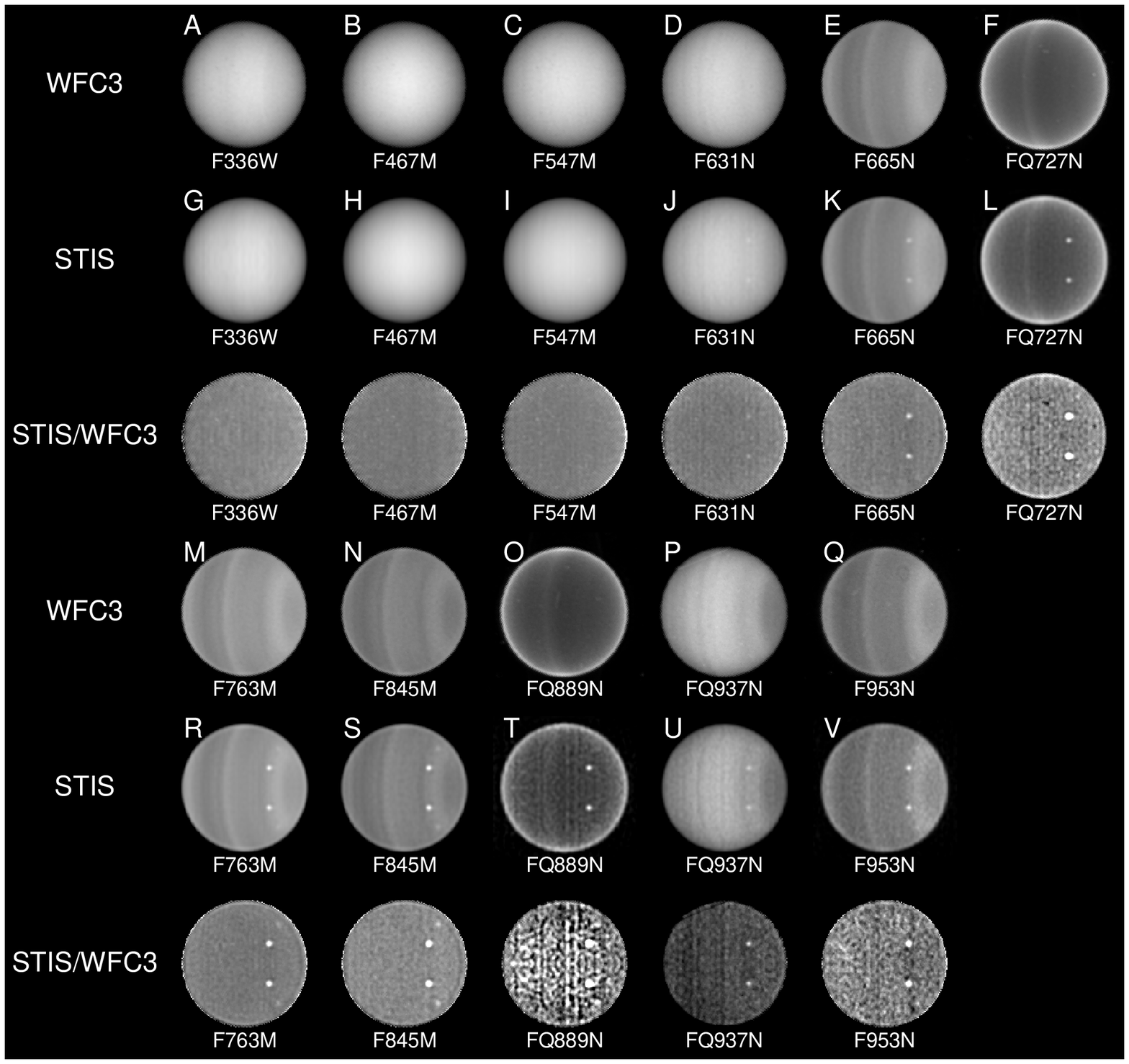}
\caption{WFC3 images of Uranus taken on 30 September 2012 (A-F and
  M-Q) compared to synthetic band-pass filter images (G-L and R-V)
  created from weighted averages of STIS spectral data cubes using
  WFC3 throughput and solar spectral weighting. The north pole is at
  at the right. Portions of the synthetic images east of the central
  meridian are obtained by reflection of the images west of the
  central meridian. That is why the single cloud feature appears twice
  in images where cloud contrast is sufficient to make the cloud
  visible. The STIS data were acquired three days before the WFC3
  data, which were taken when the discrete cloud was not visible. The
  ratio images are stretched to make 0.8 black and 1.2 white. A direct
  comparison of latitude scans at fixed view angles is provided in the
  on-line analysis supplement file.}
\label{Fig:stisimages}
\end{figure*}

Since STIS images can be radiometrically calibrated for point sources
or infinitely-extended sources, and Uranus is neither, an empirically
determined correction function must be applied to the images as a
function of wavelength. This function was determined for the 2002 STIS
observations using WFPC2 images of Uranus taken around the same time
as the 2002 STIS Uranus observations.
To ensure that this function was determined as well as possible for the
Cycle 20 observations in 2012, and to cross check the extensive
spatial and spectral corrections that are required for STIS
observations, we used one additional orbit of WFC3 imaging at a pixel
scale of 0.04 arcseconds with eleven different filters spread over the
300-1000 nm range of the STIS spectra. These WFC3 images are displayed
in Fig. \ref{Fig:stisimages}, along with synthetic images with the same
spectral weighting constructed from STIS spatially resolved spectra, as
described in the following section.
The filters and exposures
are provided
in Table\ \ref{Tbl:sci_obs}.

\section{Data reduction and calibration.}

The STIS pipeline processing used at STScI does not yield suitably
calibrated two-dimensional spectral images for an object like Uranus.
Considerable additional effort was required to reach a final calibration of
these data, using techniques developed by KT2009 and closely followed in the
calibration of the 2012 STIS observations.
Flat-fielded science images, fringe
flats, and WAVECALS from the STScI STIS data processing pipeline are
the inputs to a rather extensive post-processing suite, each step of which
is described in Section 1 of the on-line analysis supplemental file. (WAVECALS
are exposures of arc sources used for wavelength calibration.) The output
is a cube of geometrically and radiometrically calibrated
monochromatic images of Uranus.  However, this cube needs a final correction
based on comparisons with the WFC3 support images.

Each WFC3 image was deconvolved with an appropriate Point-Spread Function (PSF)
obtained from the Tiny
Tim code of \cite{Krist1995}. To match the spatial resolution of the STIS
images, the WFC3 images were then reconvolved with an
approximation of the PSF given in the analysis supplemental file.
They were then converted to I/F using header PHOTFLAM values and the
\cite{Colina1996} solar flux spectrum, averaged over the WFC3 filter
bandpasses. (The PHOTFLAM value is used to convert instrument
DN/second to flux units.) To obtain a disk-averaged I/F, the planet's
light was integrated out to 1.1 equatorial radii and averaged over the
planet's cross section in pixels, computed using NAIF ephemerides
\citep{Acton1996} and SPICELIB limb ellipse model (SPICELIB is NAIF
toolkit software used in generating navigation and ancillary
instrument information files.) The disk-averaged I/F was also computed
for each of the STIS monochromatic images, and the filter- and solar
flux-weighted I/F was computed for each of the WFC3 filter passbands
that we used.

\begin{figure}\centering
\includegraphics[width=3.2in]{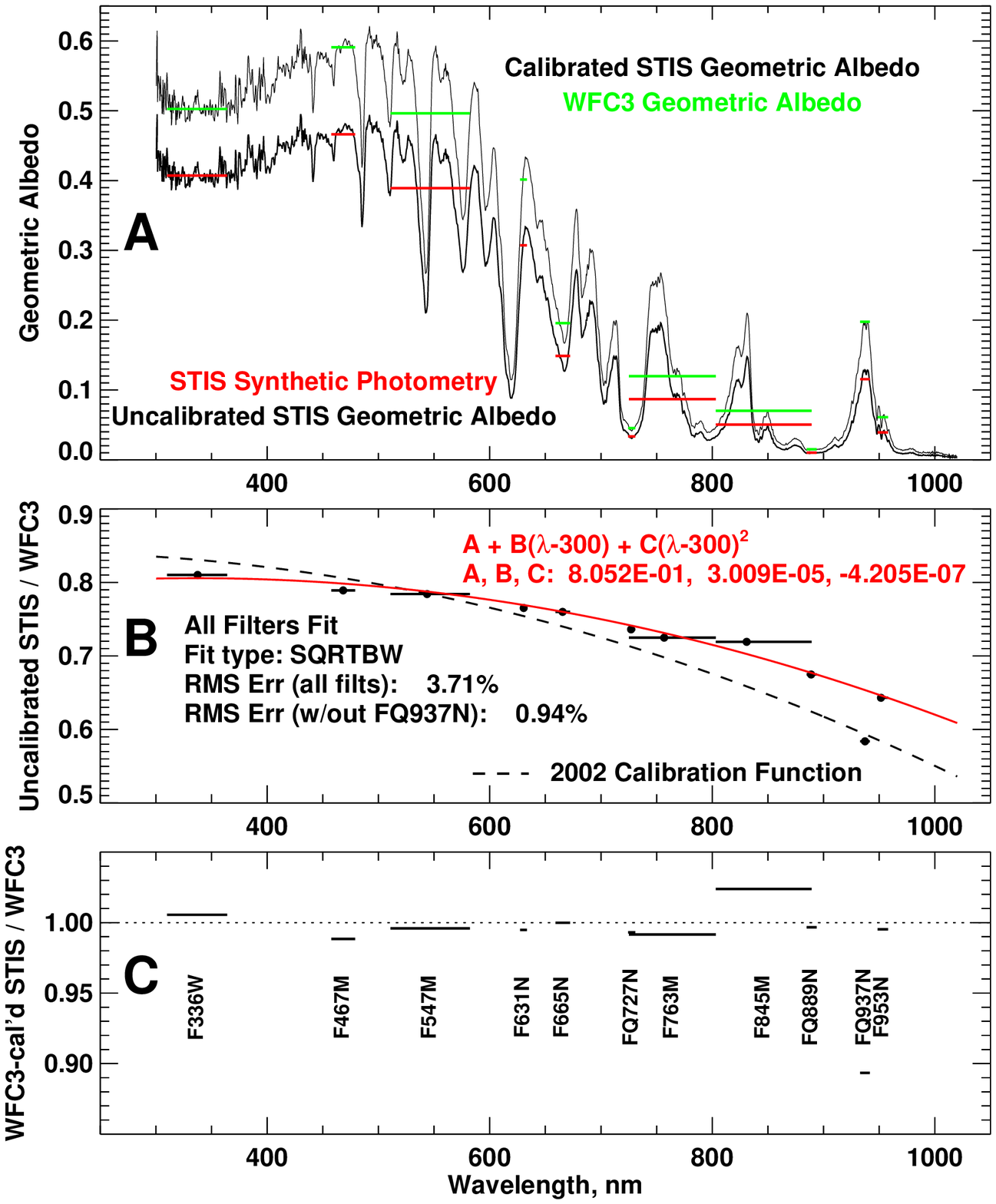}
\caption{A: Disk-integrated I/F spectra before final calibration
  (lower curve) and after final calibration (upper curve), where
  horizontal red bars indicate synthetic I/F values computed from
  pre-calibration spectra and horizontal green bars indicate I/F
  values obtained from WFC3 imaging. B: Synthetic band-pass filter
  disk-integrated I/F values (pre-calibration) divided by
  corresponding I/F values obtained from WFC3 measurements (horizontal
  bars). The red (solid) curve and legend coefficients defining it,
  were obtained by fitting all ratio values including that obtained from
  the FQ937N filter. The black (dashed) curve is the calibration
  function obtained from the 2002 data set. C: Ratio of synthetic
  disk-integrated I/F values obtained from calibrated STIS spectra to
  the corresponding WFC3 values.}
\label{Fig:pat}
\end{figure}

By comparing the uncalibrated STIS synthetic disk-averaged I/F to the
corresponding WFC3 values, we constructed a correction function to
radiometrically calibrate the STIS cube. Figure \ref{Fig:pat} shows
the ratios of STIS to WFC3 disk-integrated brightnesses, and the
quadratic function that we fit to these ratios as a function of
wavelength.  We weighted the filters by the square root of their band
widths, and computed an effective wavelength weighted according to the
product of the solar spectrum and the I/F spectrum of Uranus.  The RMS
deviation of individual filters from the calibration curve given in
Fig. \ref{Fig:pat} is 3.7\%, but only 0.94\% when the FQ937N filter is
omitted from the RMS calculation (but still included in the fit).
While the FQ937N ratio deviates by 10\% from the final calibration
curve, it is not likely due to a problem with the filter calibration,
as a comparison with the uranian satellite Ariel found it to be
consistent with the other filters.  More likely is that the darker
parts of the STIS spectrum require slightly different calibrations
than the brighter parts due to residual errors in Charge Transfer
Efficiency (CTE) corrections (described in Section 1 of the analysis
supplement file).  Figure\ \ref{Fig:pat} also shows the function that
was used to radiometrically calibrate the 2002 STIS data by KT2009.
We do not know whether the difference in functions is due to a true
STIS instrument change, to spatial variations across the CCD (2012 and
2002 observations were acquired at different locations on the CCD), or
to a difference in filters used in the two calibrations (the 2002 data
were calibrated using images taken by the Wide Field Planetary Camera
2 (WFPC2) instrument, the predecessor to WFC3).

As a sanity check on the STIS processing we compared line scans across
WFC3 images to the corresponding scans across synthetic WFC3 images
created from our calibrated STIS data cubes.  These generally were in
excellent agreement for $\mu$ values from 0.3 to 0.8, as demonstrated
by plots given in Section 2 of the on-line analysis supplement file.
The most significant discrepancy is in the overall I/F level computed
for the FQ937N filter, a consequence of our calibration curve being
10\% high for that filter.  This can also be seen
in the ratio images plotted in  Fig.\ \ref{Fig:stisimages}.

\subsection{Center-to-limb fitting}

Center-to-limb profiles provide important constraints on the vertical
distribution of cloud particles as well as the vertical variation of
methane compared to hydrogen.  Because Uranus has a high degree of
zonal uniformity, these profiles are fairly smooth functions that can
be characterized by a small number of parameters, making it possible
to constrain the profiles accurately, reducing the effects of noise
and skipping over any small discrete cloud features.  Because the
observations were taken very close to zero phase angle these functions
are almost perfectly symmetric about the central meridian, so that
they depend on only one cosine value (observer and solar zenith angles
are essentially equal). This is also one reason we were
able to characterize these profiles by measuring only one half of the
Uranus disk.  These fits also facilitate the separation of latitudinal variations
from those associated with view angle variations.

\begin{figure}\centering
\includegraphics[width=3.2in]{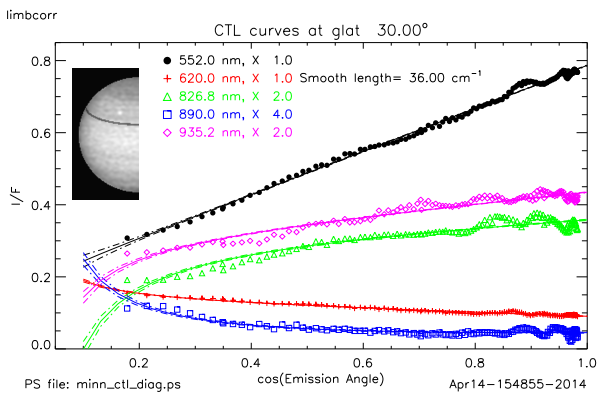}
\caption{
Sample center-to-limb fits for 30\deg N,
  as described in the main text.  STIS I/F samples and fit lines with
  uncertainty bands are shown for five different wavelengths indicated
  in the legends. The latitude band sampled for these fits is darkened
  in the inset image of the half-disk of Uranus.}
\label{Fig:ctlfits}
\end{figure}

Before fitting the center-to-limb (CTL) profile for each wavelength,
the spectral data are smoothed in the wavenumber domain to a
resolution of 36 \icmx, which equals the resolution we use in
computing the Raman spectrum, but is four times finer than the
sampling we use in constraining cloud models. The
smoothing is helpful in reducing noise at longer wavelengths without
degrading important spectral features. For each
1\deg of latitude from 54\deg S to 85\deg N, all image samples within
1\deg of the selected latitude and with $\mu > $ 0.175 are collected
and fit to the empirical function
\begin{eqnarray}
 I(\mu)= a + b \mu + c/\mu \label{Eq:ctl},
\end{eqnarray}
 assuming all samples were collected at the desired latitude and using
 the $\mu$ value for the center of each pixel of the image
 samples. Sample fits are provided in Fig.\ \ref{Fig:ctlfits} and in
Fig. 2 of the on-line supplement. Most
 of the scatter about the fitted profiles is due to noise, which is
 often amplified by the deconvolution process.  
Because the range of observed $\mu$ values decreases away from the
equator at high southern and northern latitudes, we chose a moderate
value of $\mu$=0.6 as the maximum view-angle cosine to provide a
reasonably large unextrapolated range of 33.6\deg S to 72.6\deg N for
2012 observations and 74.5\deg S to 31.7\deg N for 2002. Unless otherwise
noted all our results are derived without extrapolation.

The CTL fits can also be used to create zonally smoothed images by
replacing the observed I/F for each pixel by the fitted value. Results
of that procedure are displayed in a later section, with further examples
shown in the analysis supplement.  

\section{Direct comparison of methane and hydrogen absorptions vs. latitude.}

If methane and hydrogen absorptions had the same dependence on
pressure, then it would be simple to estimate the latitudinal
variation in their relative abundances by looking at the relative
variation in I/F values with latitude for wavelengths that produce
similar absorption at some reference latitude.  Although this idea is
compromised by different vertical variations in absorption, which
means that latitudinal variation in the vertical distribution of
aerosols can also play a role, it is nevertheless useful in a
semi-quantitative sense.  Thus we explore several cases below.

\subsection{Direct comparison of key near-IR wavelength scans in 2007}

\begin{figure}\centering
\includegraphics[width=3.2in]{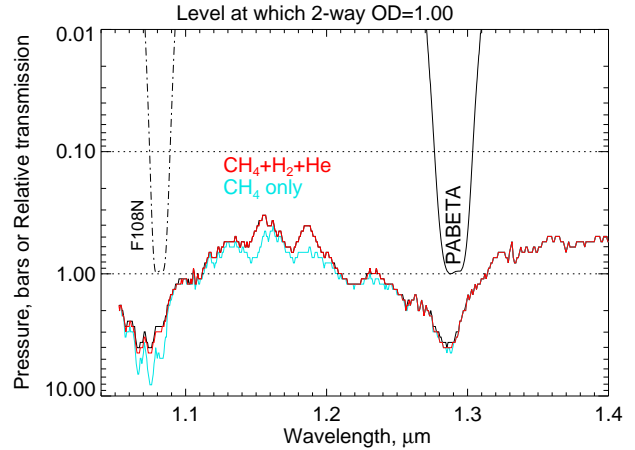}
\caption{Near-IR penetration depths vs. wavelength compared to
  filter transmission for F108N NICMOS and PaBeta Keck/NIRC2 filters,
which sense similar atmospheric levels in a clear atmosphere,
but are dominated by different gas absorptions (H$_2$ and \chf respectively).}
\label{Fig:irpendepth}
\end{figure}

\begin{figure*}\centering
\includegraphics[width=6.2in]{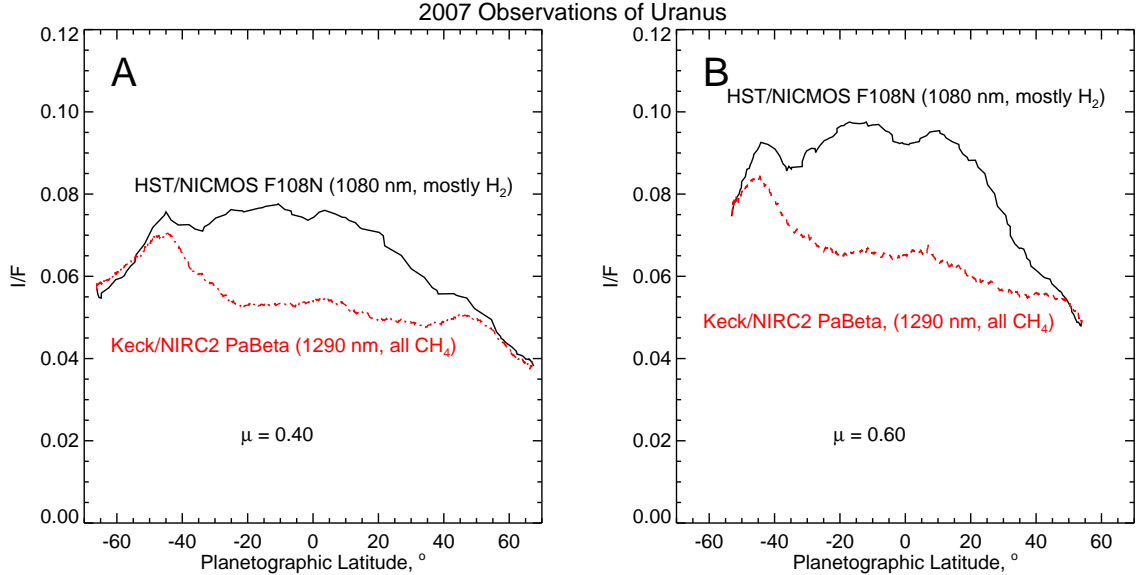}
\caption{Latitudinal profiles at fixed zenith angle cosines of 0.4 (A)
  and 0.6 (B) for F108N and PaBeta (Keck/NIRC2) filters taken near the
  Uranus equinox in 2007.  At this point the southern hemisphere was
  still generally brighter than the northern hemisphere and the 38\deg
  S - 58\deg S southern bright band was still better defined and
  considerably brighter than the corresponding northern bright band. The relatively
low equatorial I/F values for the methane-dominated PaBeta filter (1290 nm) indicates
increased CH$_4$/H$_2$ absorption at low latitudes.}
\label{Fig:nearircmp}
\end{figure*}

Our first example is based on a comparison of HST NICMOS measurements
using an F108N filter (centered at 1080 nm), which is dominated by
H$_2$ CIA, with KeckII/NIRC2 measurements using a PaBeta filter
(centered at 1290 nm), which is completely dominated by methane
absorption. [The NICMOS observation came from HST program 11118,
  L. Sromovsky PI, and was taken at 4:39 UT on 28 July 2007, at a
  phase angle of 2.00\deg and a sub-observer latitude of
  0.61\degx. The Keck II image was taken at 14:39 UT on 31 July 2007,
  at a phase angle of 1.87\deg and a sub-observer latitude of
  0.51\degx.]  That these two observations sense roughly the same
level in the atmosphere is demonstrated by the penetration depth plot
in Fig.\ \ref{Fig:irpendepth}, which also displays the filter
transmission functions. Although the Keck image has the higher spatial
resolution, both images are adequate to resolve the gross latitudinal
variations of interest.  The I/F profiles for these two filters near
the 2007 Uranus equinox are displayed in Fig.\ \ref{Fig:nearircmp} for
$\mu$=0.4 and $\mu$=0.6.  These are plotted as true I/F values (not
scaled in any way). At high latitudes in both hemispheres, and at both
zenith angle cosines, the two profiles agree with each other quite
closely and are both increasing towards the equator.  But at low
latitudes, the reflectivity profile for the hydrogen-dominated
wavelength continues to increase, while the profile for the
methane-dominated wavelength decreases substantially, indicating that
methane absorption is much higher at low latitudes than at high
latitudes.This suggests that upper tropospheric methane depletion
(relative to low latitudes) was present at both northern and southern
high latitudes in 2007, at least roughly similar to the pattern that
was inferred by \cite{Tice2013} from analysis of 2009 IRTF SpeX
observations.  Latitudinal variations in aerosol scattering could
distort these results somewhat, but because they affect both
wavelengths to similar degrees, most of the effect is likely due to
methane variations.

\subsection{Direct comparison of key STIS wavelength scans}

\begin{figure}\centering
\includegraphics[width=3.3in]{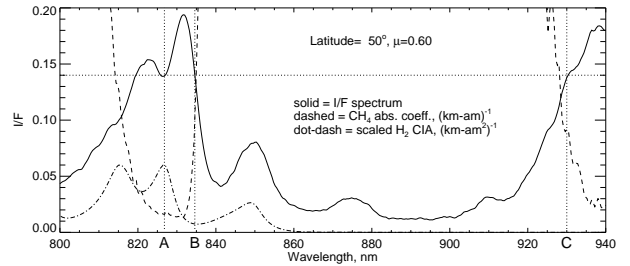}
\caption{I/F and absorption spectra comparing the equilibrium H$_2$
  CIA coefficient spectrum (divided by 1.2e-7, shown as dot-dash
  curve) and methane absorption coefficient spectrum (dashed). Note
  that the I/F spectrum has nearly equal I/F values at 826.8 nm (A),
  834.6 nm (B), and 930 nm (C), but H$_2$ absorption is much greater
  at A than at B, while the opposite is true of methane absorption,
  and at C only methane absorption is present. In a reflecting layer
  model, changes in cloud reflectivity should affect wavelengths A-C
  by the same factor, but changes in methane mixing ratio would affect
  C most and A least.}
\label{Fig:absprofiles}
\end{figure}

\begin{figure*}\centering
\includegraphics[width=6.2in]{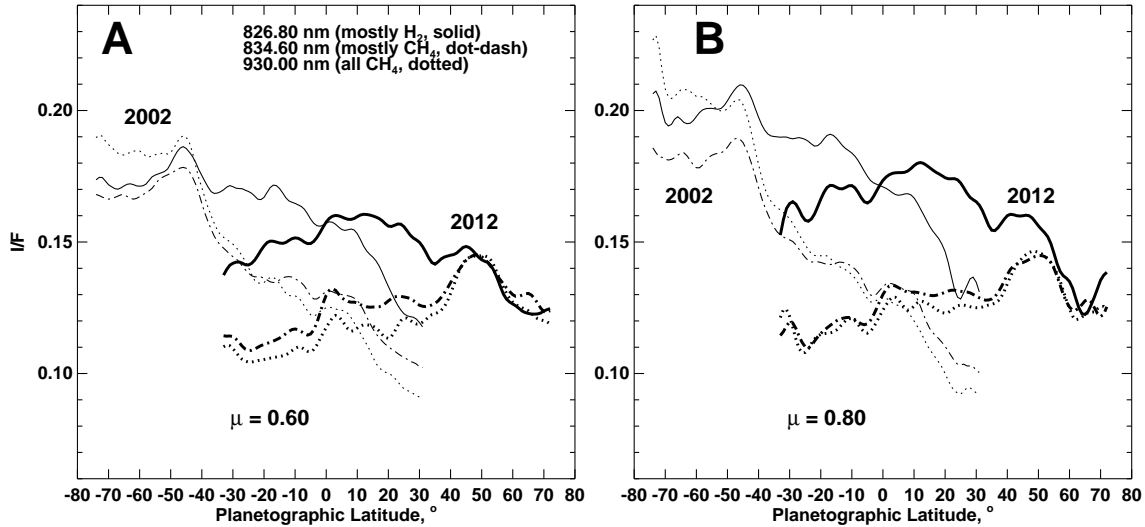}
\caption{I/F vs latitude at $\mu=0.6$ (A) and $\mu=0.8$ (B)
for three wavelengths with different amounts of methane and hydrogen absorption.
Thin curves are from 2002 and thick curves from 2012. These are plots
of center-to-limb fitted values instead of raw image data. In both cases
the methane-dominated wavelengths have much reduced I/F at low latitudes.
The 2002 calibration is based on WFPC2 comparisons and leads to I/F
values 3\% lower than the original KT2009 calibration.}
\label{Fig:latscan3}
\end{figure*}

\begin{figure*}\centering
\includegraphics[width=6in]{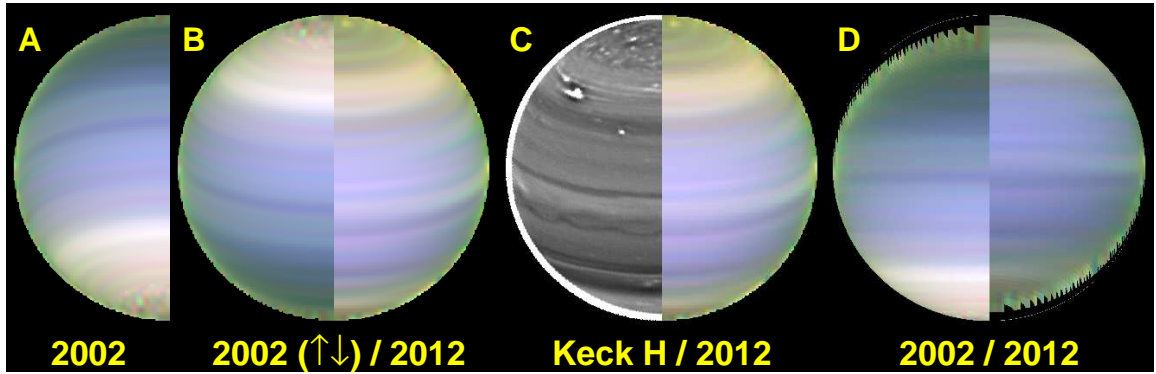}
\caption{Color composites of fitted center-to-limb smoothed images for
  2002 ({\bf A, B, D}) and 2012 ({\bf B-D}) using color assignments R
  = 930 nm (all methane), G = 834.6 nm (methane and hydrogen), and B =
  826.8 nm (mostly hydrogen).  North is up in all images except the
  reflected 2002 image in {\bf B}.  The blue tint at low latitudes for
  both years is due to locally increased methane absorption.  In {\bf
    B} the 2002 image is inverted after remapping to the same central
  latitude as the 2012 image. The obvious asymmetry is not surprising
  because Uranus was in southern fall in 2002 and northern spring in
  2012.  In {\bf C} we compare 2012 STIS observations with a
  Keck/NIRC2 near-IR image taken in the same year, high-pass filtered
  to enhance cloud structure.  Note the small cloud features north of
  the high-latitude bright band.  In {\bf D} we compare 2002 and 2012 STIS
  observations at the same latitudes (i.e. without reflection of 2002
  about the equator) and with both observations remapped to place the
  equator at the center of the image. In {\bf D} the enhancement is
  the same for both 2002 and 2012, but using the 2012 albedo
  calibration function for both data sets (the solid curve in
  Fig.\ \ref{Fig:pat}).  The southern high latitudes in 2002 were
  brighter and whiter than corresponding northern latitudes in 2012, providing evidence for a seasonal
  lag.  The dark band near the center of the two images is 5\deg south
  of the equator. A comparison of latitudinal variations at fixed view
  angles is provided in Fig.\ \ref{Fig:latscan3}.}
\label{Fig:colorcomps}
\end{figure*}

A similar spectral comparison of the 2012 STIS observations can also
be informative. By selecting wavelengths that at one latitude provide
similar I/F values but very different contributions by H$_2$ CIA and
methane, one can then make comparisons at other latitudes to see how
I/F values at the two wavelengths vary with latitude.  If aerosols did
not vary at all with latitude, then this would be a clear measure of
the ratio of \chf to H$_2$. Fig.\ \ref{Fig:absprofiles} displays a
detailed view of I/F spectral region where hydrogen CIA exceeds
methane absorption (see Fig.\ \ref{Fig:ccdpendepth} for penetration
depths). Near 930 nm and 827 nm the I/F values are similar but the
former is dominated by methane absorption and the latter by hydrogen
absorption. Near 835 nm there is a relative minimum in hydrogen
absorption, while methane absorption is still strong.  At 50\deg N
latitude and $\mu$=0.6, I/F values are nearly the same at all three
wavelengths, suggesting that they all produce roughly the same
attenuation of the vertically distributed aerosol scattering.  At low
latitudes, as shown in Fig.\ \ref{Fig:latscan3}A, the I/F for the
hydrogen-dominated wavelength increases, while the I/F for the
methane-dominated wavelength decreases substantially, indicating an
increase in the amount of methane relative to hydrogen at low
latitudes.  Similar effects are seen for the 2002 observations. For
$\mu$=0.8 (Fig.\ \ref{Fig:latscan3}B), which probes more deeply, the
changes are even more dramatic.  A color composite of these
wavelengths (using R=930 nm, G= 834.6 nm, and B= 826.8 nm) is shown in
Fig.\ \ref{Fig:colorcomps}, where the three components are balanced to
produce comparable dynamic ranges for each wavelength. This results in
nearly blue low latitudes where absorption at the two methane
dominated wavelengths is relatively high and green/orange polar
regions as a result of the decreased absorption by methane there.

The spectral comparisons in Figs.\ \ref{Fig:latscan3} and
\ref{Fig:colorcomps} also reveal substantial secular changes between
2002 and 2012. At wavelengths for which methane and/or hydrogen
absorption are important, the northern low-latitudes have brightened
substantially, while the southern low latitudes have darkened.  The
2002-2012 differences shown in Fig.\ \ref{Fig:latscan3} are not due to
a view-angle effect because comparisons in that figure are made at the
same view and illumination angles for both years. The bright bands
between 38\deg and 58\deg (north and south) have also changed
significantly, with the southern band darkening dramatically, while
the northern band brightened by a smaller amount.  The northern bright
band in 2012 was of lower contrast than the southern bright band
in 2002.  However, at continuum wavelengths (see Fig. 4 in the
analysis supplement file) secular changes are not evident. Aside from
what appears to be a small calibration disagreement between 2002 and
2012, in which corrected 2002 I/F values are about 2\% higher (at continuum
wavelengths), the latitudinal variations are very similar for both
epochs. [The corrected 2002 calibration we use here is based on WFPC2
  comparisons and leads to I/F values that are 3\% smaller than
  published by KT2009.]  The small size of these continuum differences is partly
a result of the relatively smaller impact of particulates at short
wavelengths where Rayleigh scattering is more significant.  At
absorbing wavelengths the optical depth and vertical distribution of
particulates have a greater fractional effect on I/F and thus small
secular changes in these parameters can be more easily noticed.

\vspace{0.1in}
\section{A simplified model of methane/hydrogen variations}

Here we model detailed relative latitudinal variations in effective methane
mixing ratios using a simplified fast spectral model that we put on an
absolute scale with selected full radiative transfer model results.

\subsection{Model description}

\begin{figure*}\centering
\includegraphics[width=6in]{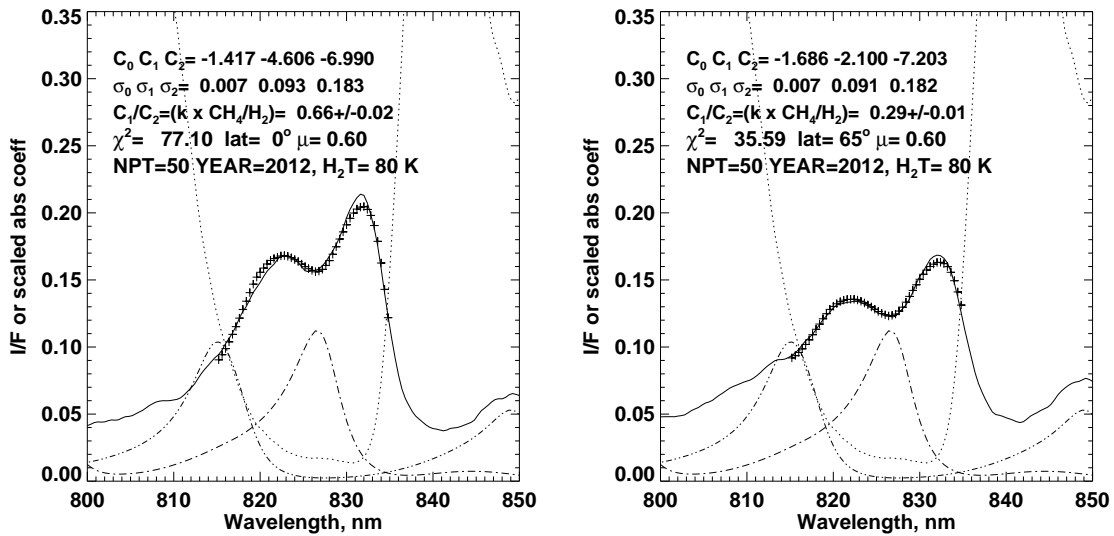}
\caption{The 2012 I/F spectrum (solid curves) from 815 to 835 nm at
  $\mu$=0.6 for two sample latitudes: 0\deg (left) and 65\deg (right),
  using the simplified model (+) of Eq.\ \ref{Eq:iofmod}, as discussed
  in the text. Also shown are relative variations of absorption by
  methane (dotted) and H$_2$ (dot-dash for para and double-dot-dash
  for ortho). }\label{Fig:simplefits}
\end{figure*}

KT2009 estimated the latitude variation of the \chf volume mixing
ratio using a simple model to fit the spectral region where hydrogen
and methane have comparable effects on the observed I/F spectrum.
They assumed that in the region from 819 nm to 835 nm the I/F spectrum
could be approximated by
\begin{eqnarray}
 I/F = \exp ( C_0 + C_1 \times k_{CH_4} + C_2 \times k_{H_2}), \label{Eq:iofmod}
\end{eqnarray}
where $k_{CH_4}$ and $k_{H_2}$ are absorption coefficients of methane
and hydrogen respectively.  While this looks like a reflecting-layer
model, it is actually more general.  The coefficients $C_1$ and $C_2$
give the weighted path lengths of observed photons. Because the methane
coefficient is essentially independent of temperature and pressure,
the weighting for coefficient $C_1$ is the methane mixing ratio. But
because $k_{H_2}$ not only depends on temperature, but also on the
square of the density, $C_2$ has a pressure-dependent weighting.  Mean
path lengths are also affected by aerosols.  The observed variation of
$C_2$ is about 10 \% peak-to-peak, aside from features that are
consistent with noise.  We expect that C$_1$ would have a similar
variation if the methane mixing ratio is constant with latitude, and
if the path lengths of photons do not vary much across the selected
spectral region.  Because $C_1$ and $C_2$ have different weightings,
one might expect a little different variation for $C_1$, perhaps 5 \%
or 15 \%.  Thus, if the ratio $C_1/C_2$ shows latitudinal variations
of about 5 \%, that could be due to aerosols.  If they are much
larger, it strongly suggests that they are due to a variation of the
methane mixing ratio.  Whether the path lengths vary across the
spectral region or not can be tested by fitting the spectral shape
with Eq. \ref{Eq:iofmod}.  The fits shown in Fig. \ref{Fig:simplefits}
are good.  In fact, the slight misfit at the peak of the curves is
likely due to a slight change of path lengths, indicating that the
chosen spectral interval cannot be expanded without introducing
systematic errors.
To evaluate the hydrogen absorption coefficient, KT2009 used an
effective fixed temperature of 80 K, which we also used because it
provided an approximate overall best fit to the spectra over a range
of latitudes.  We also followed KT2009 in assuming equilibrium
hydrogen.

While KT2009 chose ten 1.6-nm wide segments from 819 to 835 nm, we
here extended the range down to 815 nm to better sample the para
hydrogen absorption peak near that wavelength, giving us 50 wavelengths
over this range.  The disadvantage of the wider interval is that fit
quality is lower and somewhat more variable.  However, sample
calculations show very little difference in the derived latitudinal
profile.  In computing the $\chi^2$ estimates of fit quality, we
assumed a somewhat arbitrary value of 2\% in relative I/F
calibration-absorption modeling error and combined that with the
center-to-limb fitting errors using the root of the summed squares.
We used an arbitrary scaling of the $k_{H_2}$ coefficients, then
carried out a non-linear regression to obtain best fit model
coefficients, uncertainties, and the ratio of $C_1/C_2$ and its
uncertainty.  Additional uncertainty is present due to the
simplifications of the assumed model physics. The sample fits provided in
Fig.\ \ref{Fig:simplefits} show that the model is quite successful at
fitting the observed spectra, and also allows tight constraints on the
model coefficients and on the ratio of coefficients, which we use as a
proxy for the \chf/H$_2$ ratio. Section 5 of the on-line analysis
supplemental file provides a more extended discussion of this model
and our attempts to constrain both effective temperature and hydrogen
para fraction as well as relative absorber amounts.

\subsection{Model results for latitude dependence of CH$_4$/H$_2$.}

\begin{figure}\centering
\includegraphics[width=3.2in]{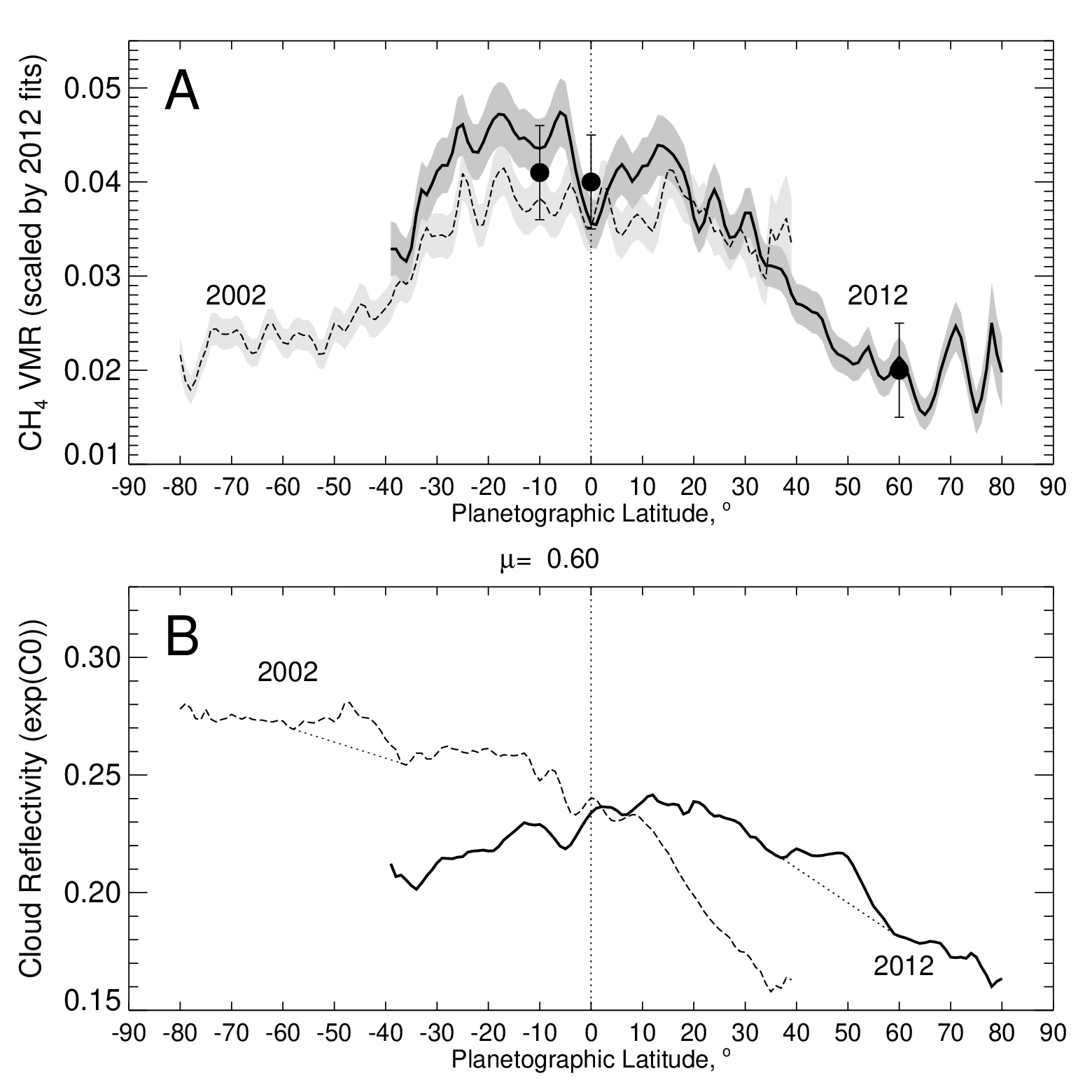}
\caption{A: Latitude dependence of the ratio $C_1/C_2$, which is
  crudely proportional to the \chfx/H$_2$ mixing ratio, scaled to best
  match the 2012 \chf VMR estimates at 10\deg S, the equator, and
  60\deg N, obtained from radiation transfer modeling. Those results
  are plotted as larger filled circles with error bars.  The same
  scaling used on the 2002 data (thin dashed line) leads to a slightly
  lower methane VMR at low latitudes. Error bars for the $C_1/C_2$
  ratio points are provided only for scattered samples for clarity. B:
  The aerosol reflectivity term (exp(C0)), for 2002 and 2012, using
  the KT2009 spectral calibration divided by 1.03 for 2002
  results. Dotted lines interpolate across the regions where bright
  bands are seen (38\deg - 58\degx).}\label{Fig:ch4latvar}
\end{figure}

Fitting this crude model to every latitude for both 2002 and 2012
center-to-limb fits leads to the latitude dependence shown in
Fig.\ \ref{Fig:ch4latvar}, where latitudes shown for each data set
include the measured ranges and slight extrapolations of about 7\deg
of latitude (the modest uncertainties of such projections are
illustrated in Fig. 2 of the on-line supplement).  The estimated error
bounds shown here, which are twice as large as the formal fitting
errors, are based on splitting the spectral data into two independent
subsets and comparing the two profile results.  Here we use a scaling
of the $C_1/C_2$ ratio that best matches the methane volume mixing
ratio (VMR) values estimated from full radiation transfer modeling
that properly accounts for the density dependence and temperature
dependence of hydrogen absorption (discussed in subsequent sections).
It is noteworthy that a single scale factor applied to all latitudes
yields consistency with the full radiation transfer results at three
different latitudes with different aerosol reflectivities.  Using the
same scale factor on the 2002 profile leads to about a 4\% overall
average between 30\deg S and 20\deg N.  There are substantial
differences between 2002 and 2012 where the two profiles overlap,
though many of the small scale variations occur at nearly the same
latitudes. The relative variations between 35\deg S and 35\deg N have
a correlation of 70\%, with the 2012 variations being about 10\%
larger than the 2002 variation. Yet the average 2012 methane VMR is
closer to 4.5\% at low southern latitudes where the 2002 profile is
closer to 3.8\%.  Some of this difference might be due to the
difference in aerosol structure, which is indicated by the higher
reflectivity of the southern hemisphere in 2002 (see
Fig.\ \ref{Fig:latscan3}). Most of the small scale variability seen in
the 2002 ratio is due to the hydrogen absorption term. Because this
occurs without any evident small scale variability in the aerosol term
(Fig.\ \ref{Fig:latscan3}B), it suggests the possibility that
small-scale temperature variations or vertical mixing variations that
produce changes in para fractions might be responsible (this is
discussed further in the next paragraph).  A significant part of the
small scale structure seen in the 2012 methane VMR values is also due
to hydrogen coefficient variations, especially north of 60\deg N where
the model cloud reflectivity varies relatively smoothly with latitude
(Fig. \ref{Fig:ch4latvar}B), at least at these wavelengths (small
variations are seen at near-IR wavelengths, as discussed later).

Also noteworthy is the difference between 2012 north and 2002 south
polar regions (there is no overlap between 2002 and 2012 in either
polar region).  In 2002 the mixing ratio from 75\deg S to 50\deg S is
close to 2.4\% and exhibits only small variations with latitude, while
the 2012 profile at equivalent northern latitudes displays substantial
and nearly sinusoidal variations with a latitudinal wavelength near
10\degx. It should be noted that each of these points is derived from
50 wavelengths and is thus far less noisy than one might suspect based
on the plots of single wavelengths, as in Figs.\ \ref{Fig:latscan3}
and \ref{Fig:colorcomps}.  The reality of these high-latitude features
is further supported by the fact that the same analysis applied to
similar latitudes in the opposite hemisphere using 2002 observations
did not find such large variations, suggesting that the variations are
more likely due to a hemispheric difference rather than an inherent
high latitude artifact of the analysis.  Also, there is no obvious
reason why such artifacts should suddenly disappear at latitudes below
60\deg N.  Further evidence supporting the reality of these
latitudinal variations in the methane to hydrogen absorption ratio is
the discovery of a correlated pattern of cloud reflectivity variations
in high-resolution Keck imagery, which is discussed in the analysis
supplement file.  A hint of this pattern can also be seen in
Fig.\ \ref{Fig:colorcomps}C.  We found similar patterns in August and
November images and both had moderately significant negative
correlations with the methane VMR ($\sim$ 9 \% for August and($\sim$ 5-10 \% for
November).  While the correlations are suggestive, it is still possible
that the STIS latitudinal variation is due to noise and new observations
will be needed to confirm this feature.  The explanation for
this feature, presuming that it is real,  might be that latitude bands of reduced
above-cloud methane lead to clouds appearing locally brighter due to
reduced methane absorption.  An alternate explanation is that the
local variation in methane to hydrogen absorption is due to changes in
the para fraction of hydrogen.  If we force the methane mixing ratio
to be constant over the 55\deg- 82\deg latitude range and fit spectral
variations by adjusting the para fraction and effective temperature,
then para fraction increases appear where we previously had methane
decreases.  The para increases suggest downwelling and thus we have a
correlation of descending motion with reduced cloud amount (or reduced
cloud reflectivity).  This alternative interpretation presents a more
appealing picture than our initial interpretation, although it has
a slightly worse spectral fit quality.

\section{Full radiative transfer modeling of methane and aerosol distributions}

The spectral profile comparisons (Figs.\ \ref{Fig:nearircmp},
\ref{Fig:absprofiles}, and \ref{Fig:latscan3}), and especially the
somewhat more quantitative modeling of Fig.\ \ref{Fig:ch4latvar},
strongly suggest that there is a permanent and more or less symmetric
high latitude methane depletion on Uranus.  But, because hydrogen
absorption has a density squared dependence, and the methane has a
vertically varying mixing ratio, a more accurate constraint on methane
requires full radiative transfer modeling, including the effects of
more realistic aerosol distributions, as described in the following.
The more complete modeling is also needed to provide the scaling
factor that converts the $C_1/C_2$ ratio plotted in
Fig. \ref{Fig:ch4latvar} to a methane VMR.

\subsection{Radiation transfer calculations}\label{Sec:radtran}
We used the radiation transfer code described by \cite{Sro2005raman},
which include Raman scattering and polarization effects on outgoing
intensity, though this is a minor virtue at the wavelengths employed
in our analysis.  To save computational time we employed the accurate
polarization correction described by \cite{Sro2005pol}.  After trial
calculations to determine the effect of different quadrature schemes
on the computed spectra, we selected 14 zenith angle quadrature points
per hemisphere and 14 azimuth angles. Calculations with 10 quadrature
points and 10 azimuth angles changed fit parameters by only about 1\%, which is much less than
their uncertainties.  To characterize methane absorption at CCD
wavelengths we used the coefficients of KT2009.
To model collision-induced opacity of H$_2$-H$_2$ and He-H$_2$
interactions, we interpolated tables of absorption coefficients as a
function of pressure and temperature that were computed with a program
provided by Alexandra Borysow \citep{Borysow2000}, and available at
the Atmospheres Node of NASA'S Planetary Data System. We assumed
equilibrium hydrogen for most calculations, following
KT2009 and \cite{Sro2011occult}.

\subsection{Cloud model}

\begin{figure}\centering
\includegraphics[width=3.2in]{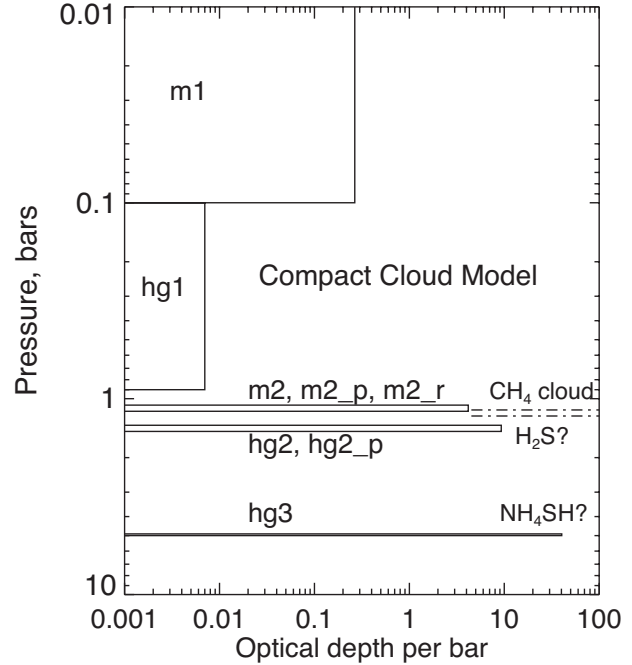}
\caption{Compact cloud model vertical structure. Compared to the
  KT2009 model, this has two layers with additional adjustable parameters
  of base pressure (for the new $hg2$ layer) and base pressure and
  particle radius (for the new $m2$ layer).}
\label{Fig:structuremodels}
\end{figure}

The cloud model we used (referred to as the compact model in what
follows) is a modification of the KT2009 diffuse model (containing 4
vertically extended layers) that is fully described in Section 7 of
the supplemental analysis file.  Our five-layer model is illustrated in
Fig.\ \ref{Fig:structuremodels} and summarized in Table
\ref{Tbl:compactmodel}. In this table, parameter names begin with
$m$ for layers containing Mie particles and with $hg$ for layers containing particles
with double Henyey-Greenstein phase functions. Here we use Mie particle
scattering as a convenient parameterization of wavelength dependence, realizing
that the particles are generally not spherical. Adjustable
parameters are identified in the Value column of the table.  We chose this model because it provided
substantially better fits than the diffuse model.  Our top two
haze layers use the same parameterization as KT2009.  The main change we
made to the KT2009 model was to replace its middle tropospheric
layer with two compact layers: an upper middle tropospheric cloud
layer (UMTC) and a lower middle tropospheric cloud layer (LMTC). The
UMTC layer is composed of Mie particles, which we characterized by a
gamma size distribution with an adjustable mean particle radius and a
fixed normalized variance of 0.1, a fixed refractive index of 1.4, and
an imaginary index of zero. 
For the LMTC layer we used particles with the same scattering
properties as given by KT2009 for their tropospheric particles.  Both
of these compact layers have the bottom pressure as a free
(adjustable) parameter and a top pressure that is a fixed fraction of
0.93 $\times$ the bottom pressure.  This degree of confinement is
approximately the same as obtained for the cloud layer inferred from
the radio occultation analysis \citep{Sro2011occult}. The motivation for
introducing these replacement layers was to obtain more flexibility in
vertical structure and to allow the possibility of including a thin
cloud near the methane condensation level, as suggested by the
occultation analysis.

\begin{table*}\centering
\caption{Compact cloud model parameters}
\vspace{0.15in}
\begin{tabular}{l l l }
\hline
                     &  Parameter/function                  &  Value        \\ 
\hline
Stratospheric haze  &  $m1\_pd$ (bottom pressure)   & fixed at 100 mb \\
of Mie particles    &  $m1\_r$ (particle radius)    & fixed at 0.1 \mum \\
with gamma size      &  $m1\_b$ (variance)           & fixed at 0.3 \\
distribution         &  $n1$ (refractive index)      & nr=1.4, ni given by KT2009\\
                     &  $m1\_odpb$ (optical depth/bar) & adjustable1 \\
\hline
Upper tropospheric haze & $hg1\_pd$ (bottom pressure)  & fixed at 0.9 bars (top pressure = $m1\_pd$) \\
of double HG particles  & $\varpi (\lambda)$ (single-scatt. albedo) & KT2009 function\\
(UTH)                        & phase function (KT2009) & $g_1=0.7$, $g_2=0.3$, $f_1(\lambda)$ given by KT2009\\
                        & $hg1\_odpb$ (optical depth/bar) & adjustable2\\
\hline
Upper middle tropospheric & $m2\_p$ (bottom pressure)  & adjustable3 (top pressure = $m2\_p \times$0.93) \\
cloud layer of Mie  &  $m2\_r$ (particle radius)    & adjustable4 \\
particles (UMTC)        &   $m2\_b$ (variance)           & fixed at 0.3 \\
                        &  $n2$ (refractive index)      & fixed at n=1.4\\
                        &  $m2\_od$ (optical depth) & adjustable5 \\
\hline
Lower middle tropospheric & $hg2\_p$ (bottom pressure) & adjustable6 (Ptop = $hg2\_p$ X 0.9)\\
cloud of double HG  &  $\varpi (\lambda)$ (single-scatt. albedo) & given by KT2009\\
particles (LMTC)       &  phase function (KT2009) & $g_1=0.7$, $g_2=0.3$, $f_1(\lambda)$ given by KT2009\\
                        & $hg2\_od$ (optical depth) & adjustable7\\
\hline
Bottom tropospheric cloud &  $\varpi (\lambda)$  & same as previous layer\\
  (BTC)                 & phase function (DHG) & same as previous layer\\
                        & $hg3\_od$ (optical depth) & adjustable8\\
                        & $hg3\_p$ (bottom pressure) & fixed at 5 bars\\ 
\hline\\
\end{tabular}\label{Tbl:compactmodel}
\vspace{-0.1in}
\parbox[]{5.5in}{NOTE: KT2009 equations defining wavelength dependent
  parameters are reproduced in the analysis supplement file. Usually
  $hg1\_odpb$ was found to be too small to bother including in our
  fits.}
\end{table*}

The last change we made was to replace the KT2009 bottom tropospheric
layer by a compact cloud layer at 5 bars (the BTC, or bottom
tropospheric cloud), with adjustable optical depth and with the KT2009
tropospheric scattering properties. \cite{Sro2011occult} found that
this layer was needed to provide accurate fits near 0.56 and 0.59
\mumx, but its pressure was not well constrained by the observations
(pressure changes could be compensated by optical depth changes, to
produce essentially the same fit quality). Whether this deep cloud is
vertically diffuse or compact also could not be well constrained.
Sample phase functions and the scattering properties of particles used
in our cloud model are plotted vs. wavelength in the analysis
supplement file (Section 7). Note that the KT2009 empirical functions
defining wavelength-dependent single-scattering albedo and phase
functions only apply over the 300 nm to 1000 nm range for which they
were derived.  The role played by each model layer in creating a
spectrum that matches sample observations is also provided in the
analysis supplement file (Section 8).

\subsection{Fitting procedures}

To avoid the complexity of fitting a wavelength-dependent imaginary index in the methane layer
(the UMTC layer in Table \ref{Tbl:compactmodel}) we fit only the
wavelength range from 0.55 \mum to 1.0 \mumx.  We chose a wavenumber
step of 118.86 \icm for sampling the observed and calculated 
spectrum. This yielded 69 spectral samples, each at three different
zenith angle cosines (0.3, 0.4, and 0.6), for a total of 207 points of
comparison.
Our compact layer model has seven to eight adjustable parameters (see
Table\ \ref{Tbl:compactmodel}),
leaving 201 or 200 degrees of freedom ($N_F$) respectively. (The
expected value of \chisq is equal to $N_F$ for an optimized fit, and
the formal uncertainty in any fit parameter is given by the change in
its value that increases \chisq by one unit when the remaining
parameters are optimized \citep{Press1992}.) We fixed the BTC base
pressure to 5 bars (see Table \ref{Tbl:compactmodel}) and usually
ignored the UTH layer because of its negligible effects.
We used a modified Levenberg-Marquardt non-linear fitting algorithm
\citep{Sro2010iso} to adjust the fitted parameters to minimize
$\chi^2$ and to estimate uncertainties in the fitted
parameters. Evaluation of \chisq requires an estimate of the expected
difference between a model and the observations due to the
uncertainties in both. We followed the same approach for estimating
uncertainties as used by \cite{Sro2011occult}, which combined
measurement noise (estimated from comparison of individual
measurements with smoothed values), modeling errors of 1\%, relative
calibration errors of 1\% (larger absolute calibration errors were
treated as scale factors), and effects of methane absorption
coefficient errors, taken to be random with RMS value of 2\% plus an
offset uncertainty of 5$\times10^{-4}$ (km-amagat)$^{-1}$.  The
uncertainty in $\chi^2$ is expected to be $\sim$25, and thus fit
differences within this range are not of significantly different
quality.

\section{Compact model fits}

The compact model described previously provides generally
better fits than the diffuse models, reaching \chisq values in the low to mid
200's compared to mid 300's to low 500's for the diffuse model. This
is presumably due to its greater flexibility to fit vertical
distributions because the main cloud layer is divided into two
sublayers with adjustable pressures as well as adjustable optical
depths.  Another difference from diffuse models is that its upper
sublayer has its wavelength dependence controlled by Mie scattering,
which is parameterized by particle radius and refractive index, rather
than the wavelength-dependent double Henyey-Greenstein parameters of
of the lower sublayer, which we take to be those of KT2009 and reproduced
in the analysis supplement file.

In the following, we first constrain the effective methane mixing
ratio profile at key latitudes to define the scaling factor for
Fig.\ \ref{Fig:ch4latvar}, then try to fit cloud structure to the
observed spectra over a wide range of latitudes, using the best-fit
equatorial methane vertical profile to discover where it does not
apply. Poleward of 30\deg N we obtain improved fits by
reducing the effective mixing ratio.  But the best fits are obtained
with a depletion profile that exhibits decreasing depletion with
increasing depth, which also is more physically plausible than a
uniform depletion at all depths.  In other words, the preferred latitudinal
profile of methane becomes more uniform at greater depths.

\subsection{Constraining equivalent methane mixing ratios at key latitudes}

\begin{figure}\centering
\includegraphics[width=3.2in]{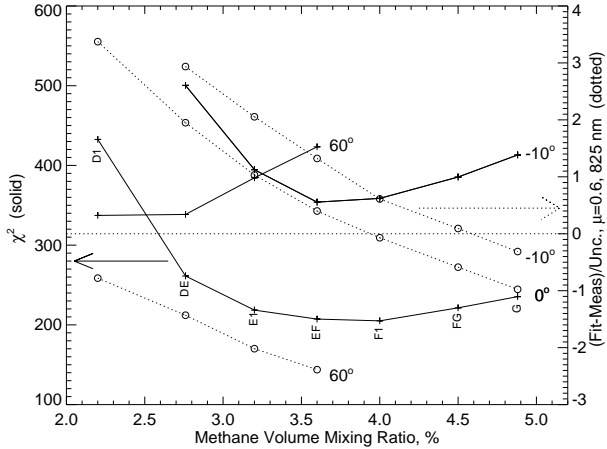}
\caption{Best-fit compact model parameters vs deep methane mixing
  ratio for undepleted occultation-consistent D1, E1, EF, F1, and G
  profiles from \cite{Sro2011occult}, and DE and FG profiles created using the
  same procedure. The fits are to STIS spectra at 10\deg S, the
  equator, and 60\deg N, for zenith angle cosines of 0.3, 0.4, and
  0.6.  The \chisq values are shown with solid curves, and the 825-nm
  error ratio to expected error at $\mu$=0.6 is shown with dotted curves. Horizontal
arrows indicate which axis to read in each case.}
\label{Fig:compact_vs_ch4_lowlat}
\end{figure}

The detailed latitudinal variation of the equivalent methane mixing
ratios plotted in Fig.\ \ref{Fig:ch4latvar} is based on a scaled ratio
of fit coefficients ($C_1/C_2$).  To define that scaling, we used fits
that fully account for the different vertical absorption profiles of
hydrogen CIA and methane, carried out at three latitudes covering the
range of mixing ratio variation (10\deg S, 0\degx, and 60\deg N).  To
estimate the optimum equivalent deep methane mixing ratio at each
latitude we did compact model fits for a variety of methane profiles
(D1, DE, E1, EF, F1, FG, and G) that have a range of deep methane
mixing ratios (2.22\%, 2.76\%, 3.20\%, 3.6\%, 4.00\%, 4.5\%, and
4.88\%, respectively). These profiles are all consistent with the
Voyager 2 occultation measurements of \cite{Lindal1987}, and thus also
have different temperature and above-cloud methane profiles. They are
all from \cite{Sro2011occult} except for DE and FG, which we 
constructed to be occultation consistent using the same techniques as
\cite{Sro2011occult}.  The fit results at our selected key latitudes
are plotted in Fig.\ \ref{Fig:compact_vs_ch4_lowlat}, where we show
two fit quality measures as a function of the deep methane mixing
ratio of the profiles that were employed.  The two measures are (1) \chisq
and (2) the signed error at 825 nm divided by the error expected just from
measurement errors.  At 10\deg S the minimum \chisq occurs at a
lower methane mixing ratio (3.6\%) than the minimum error at 825 nm (4.6\%),
while at the equator the two measures agree on 4\%. 
For 60\deg N, we find by extrapolating that the 825-nm error
reaches zero at 1.6\%, while the corresponding \chisq minimum is
between 2.2\% and 2.6\%. 
We consider the 825-nm estimate to be the most relevant since it
provides the most direct comparison between methane and hydrogen
absorptions and uses the $\mu$ = 0.6 spectrum to weight the deep
atmosphere more than the overall fit, which uses $\mu$ = 0.3, 0.4, and
0.6. However, its greater uncertainty limits led us to use
intermediate estimates between these two measures: 4.1$\pm$0.5\% at 10\deg S,
4.0$\pm$0.5\% at the equator, and 2.0$\pm$0.5\% at 60\deg N.  These
points were plotted in Fig.\ \ref{Fig:ch4latvar} as filled circles.
That these results lead to such a consistent scaling for the
simplified model justifies its use in estimating latitudinal
variations.  The low-latitude average mixing ratio is sufficiently
close to the F1 profile value of 4\% that we used F1 as our base
profile.

Note that an occultation consistent profile with a
deep mixing ratio as low as 1.6\%, which is the deep VMR assumed by
\cite{Tice2013}, would be grossly inconsistent with the observed
spectra at low latitudes.  Though \cite{Tice2013} were able to fit
their spectra with such low methane VMR values, they may have been
able to make up for lacking methane absorption at longer wavelengths
by assuming a low single-scattering albedo in their main cloud
layer.  Also, based on sample spectral fits shown in their Figs. 4, 7,
and 15, they did not fit very well the critical spectral region near
825 nm, where their model spectra show relatively more hydrogen
absorption than the observations, which suggests that their methane
mixing ratio is too low. They also reduce the weight of this spectral
region in their fits by assigning relatively large uncertainties
compared to other regions with similar I/F values.  Their fits were
also aided by using much more above-cloud methane than would be
allowed by the 0\% relative humidity implied by the Voyager 2
occultation results of \cite{Lindal1987} for their assumed deep mixing
ratio.

\subsection{Compact model fits versus latitude for the F1 profile}

\begin{figure*}\centering
\includegraphics[width=3.1in]{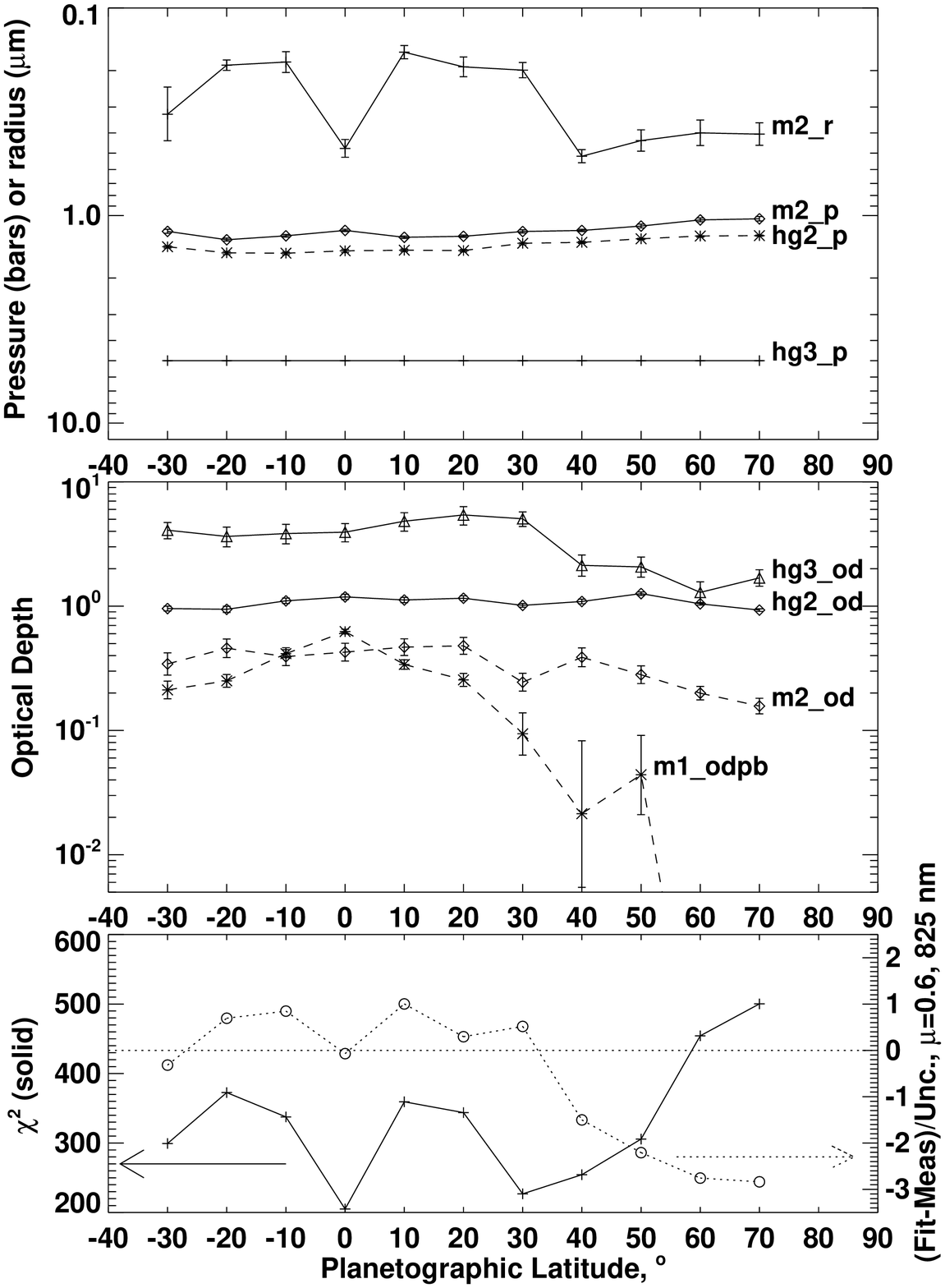}
\includegraphics[width=3.1in]{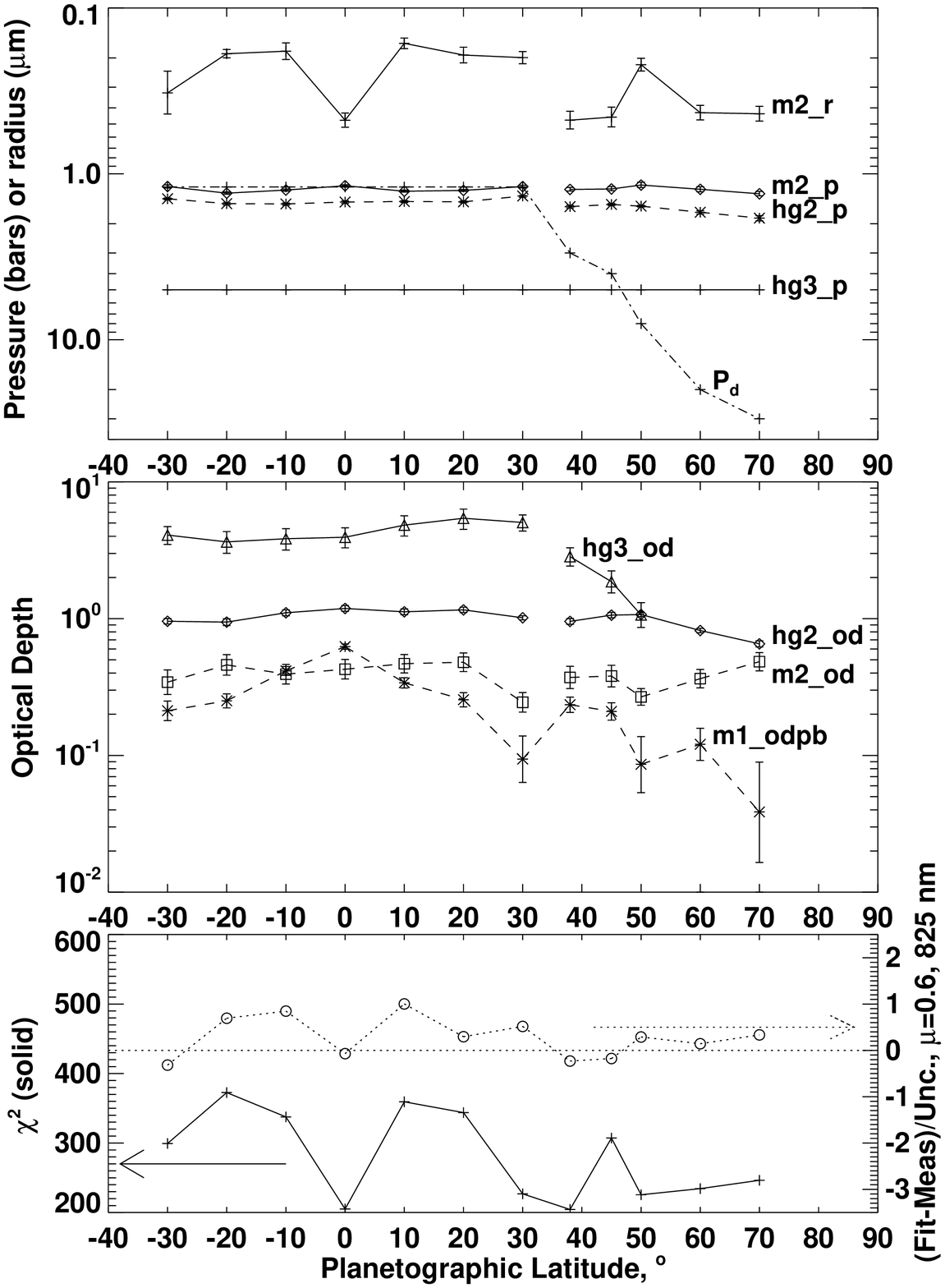}
\caption{Best-fit compact model parameters vs. latitude assuming the
  F1 structure and methane mixing ratio profile (left) and using the
  best-fit depletion depth profiles (right). Error bars indicate
  fitting uncertainty.  In each case, the lower panel displays fit
  quality in terms of \chisq (+) and signed 825-nm error (o).  In the
  bottom left plot, the behavior of the 825-nm error indicates that
  the actual methane absorption declines below the assumed value
  beginning about 40\deg N.  In the bottom right plot, both 825-nm
  error and \chisq benefit from use of depleted profiles. In the bottom
panels horizontal arrows indicate which axis to read for each curve.}
\label{Fig:compactvslat}
\end{figure*}

 We first tried to fit the compact model to a range of latitudes,
 while assuming that the F1 profile of temperature and methane applied
 to all latitudes.  The entire set of results is plotted in
 Fig.\ \ref{Fig:compactvslat}, and results for latitudes up to 30\deg
 N are presented in Table\ \ref{Tbl:compactF1simvclat}.
There are several remarkable features of these fits.  First, the $m1$
vertically diffuse stratospheric haze layer is found to have
significantly reduced opacity north of 25\deg N.  Whether this is a real effect
and a real change from the structure inferred from 2002 observations or a
result of a different spectral calibration using WFC3 observations is
unknown.  Perhaps a related difference is that the
upper sublayer (UMTC) of the main cloud layer, a Mie layer, has a fitted
particle radius generally in the 0.2-0.5 micron range, compared to
the 1.2-micron radius that seemed to be preferred by the \cite{Sro2011occult}
analysis of 2002 STIS observations.  This is
likely due to the revised albedo calibration function
(Fig.\ \ref{Fig:pat}), which produces a stronger I/F decline with
wavelength that is better matched by smaller particles (see analysis
supplement file).

\begin{table*}\centering
\caption{Best-fit parameters for equatorial compact cloud layer models
  versus latitude assuming simulated F1 profile for latitudes between
  30\deg S and 30\deg N, and using depleted F1 profiles for higher
  latitudes.}
\vspace{0.15in}
\begin{tabular}{r c c c c c c c c c}
\hline
Lat.  & $m2\_p$     & $hg2\_p$     & $m1\_odpb$   & $m2\_od$     & $hg2\_od$    & $hg3\_od$    & $m2\_r$     & $\chi^2$   &(m-o)/u\\
  \deg &  bar      &  bar      &            &            &            &            & \mum &      \\[0.05in]
\hline\\[-0.1in]
-30 &  1.19$\pm$0.03 &  1.41$\pm$0.02 &  0.21$\pm$0.04 &  0.34$\pm$0.09 &  0.96$\pm$0.05 &  4.09$\pm$0.75 &  0.32$\pm$0.12 &  299 &  -0.32\\[0.05in]
-20 &  1.31$\pm$0.03 &  1.51$\pm$0.04 &  0.25$\pm$0.04 &  0.46$\pm$0.11 &  0.94$\pm$0.07 &  3.64$\pm$0.90 &  0.19$\pm$0.01 &  373 &   0.69\\[0.05in]
-10 &  1.25$\pm$0.02 &  1.52$\pm$0.03 &  0.42$\pm$0.04 &  0.39$\pm$0.09 &  1.10$\pm$0.07 &  3.84$\pm$0.90 &  0.18$\pm$0.03 &  338 &   0.85\\[0.05in]
  0 &  1.18$\pm$0.02 &  1.48$\pm$0.03 &  0.62$\pm$0.03 &  0.43$\pm$0.07 &  1.19$\pm$0.06 &  3.94$\pm$0.67 &  0.47$\pm$0.05 &  205 &  -0.07\\[0.05in]
 10 &  1.27$\pm$0.02 &  1.47$\pm$0.03 &  0.34$\pm$0.04 &  0.47$\pm$0.10 &  1.12$\pm$0.07 &  4.83$\pm$1.09 &  0.16$\pm$0.02 &  359 &   1.00\\[0.05in]
 20 &  1.26$\pm$0.02 &  1.47$\pm$0.03 &  0.26$\pm$0.04 &  0.48$\pm$0.10 &  1.16$\pm$0.07 &  5.42$\pm$1.20 &  0.19$\pm$0.03 &  344 &   0.29\\[0.05in]
 30 &  1.19$\pm$0.02 &  1.36$\pm$0.02 &  0.09$\pm$0.04 &  0.24$\pm$0.04 &  1.01$\pm$0.04 &  5.05$\pm$0.72 &  0.20$\pm$0.02 &  227 &   0.52\\[0.05in]
 38 &  1.24$\pm$0.03 &  1.58$\pm$0.03 &  0.23$\pm$0.03 &  0.37$\pm$0.07 &  0.95$\pm$0.05 &  2.84$\pm$0.44 &  0.47$\pm$0.06 &  204 &  -0.23\\[0.05in]
 45 &  1.23$\pm$0.04 &  1.53$\pm$0.03 &  0.21$\pm$0.04 &  0.38$\pm$0.09 &  1.06$\pm$0.06 &  1.86$\pm$0.43 &  0.45$\pm$0.08 &  307 &  -0.18\\[0.05in]
 50 &  1.17$\pm$0.05 &  1.57$\pm$0.02 &  0.09$\pm$0.05 &  0.27$\pm$0.04 &  1.07$\pm$0.04 &  1.07$\pm$0.24 &  0.22$\pm$0.02 &  226 &   0.29\\[0.05in]
 60 &  1.24$\pm$0.05 &  1.70$\pm$0.04 &  0.12$\pm$0.04 &  0.36$\pm$0.06 &  0.82$\pm$0.03 &  0 &  0.43$\pm$0.05 &  234 &   0.14\\[0.05in]
 70 &  1.32$\pm$0.05 &  1.85$\pm$0.07 &  0.04$\pm$0.04 &  0.48$\pm$0.08 &  0.65$\pm$0.04 &  0 &  0.43$\pm$0.05 &  246 &   0.33\\[0.05in]
\hline
\hline\\[-0.15in]
\end{tabular}\label{Tbl:compactF1simvclat}
\parbox{5.6in}{Note: The fits from -30\deg to 30\deg were done using the F1
  thermal and methane profile, which has a deep methane volume mixing ratio of 4\%, and a
  He/H$_2$ ratio of 0.1306. At other latitudes the fits used methane depleted
profiles with $vx$=3.0 and depletion depths between the value that minimizes 825-nm errors
and those that minimize \chisq (both listed
in Table\ \ref{Tbl:pdvslat} and plotted in Fig.\ \ref{Fig:compactvslat}).
 The uncertainty in \chisq is $\sim$25 and thus fits differing by less
than this are not of significantly different quality. The column
labeled $(m-o)/u$ is the fit error at 0.825 \mum expressed as the
ratio of (model I/F - observed I/F) to the estimated I/F uncertainty
for $\mu$ = 0.6. These results are plotted in
Fig.\ \ref{Fig:compactvslat}. The parameter $hg1\_odpb$ was omitted
from these fits for reasons discussed in the main text.}
\end{table*}

At the equator the results are also unusual. In diffuse model fits we
found that the upper tropospheric haze (the second layer in Table
\ref{Tbl:compactmodel}) provided a sharply increased contribution at
the equator, and seemed to be the main change associated with the
bright equatorial band visible in many images at wavelengths of
intermediate methane absorption (e.g. Fig.\ \ref{Fig:stisimages}K, and
S).  In the compact model fits, however, changes in the next lower and
higher layers (first and third layers in Table \ref{Tbl:compactmodel})
seem to be relatively more important, with changes in pressure and
particle size of the third layer ($m2\_p$ and $m2\_r$) as well as the
optical depth per bar of the top layer ($m1\_odpb$) playing a
role. The derived value of the optical depth per bar of the second
layer ($hg1\_odpb$) for the compact model is comparable to its
uncertainty and was ignored for most compact model fits.

Between 30\deg S and 30\deg N, the 825-nm signed error, given by
(model-observed)/uncertainty, is relatively flat. 
The small positive error seen in this at most
latitudes in this range suggests that the mixing ratio in the F1
profile is not quite large enough. The rather large negative 825-nm
errors found at high latitudes are associated with excessively high
methane mixing ratios in the model calculations.  The \chisq measure
of fit quality is generally low, but is a more erratic function of latitude over this range.
Beyond 40\deg N, the 825-nm signed error becomes much larger as the
overall fit quality measured by \chisq becomes much worse.
This indicates that the assumed latitude-independent profile of
methane is not consistent with the observations.  A different methane
profile is clearly needed at high latitudes as was already
demonstrated at 60\deg N from fits used to define the scaling of the
simplified model.

\subsection{Why methane depletion should not extend to great depths}

While structure and methane profiles with a 2\% \chf deep mixing ratio
provide the best fits at 60\deg N compared to other profiles with
constant mixing ratios up to the condensation level, the deep contrast
between equator and pole (2\% VMR to 4\% VMR) is not physically
plausible.  Such deep latitudinal gradients in composition would lead
to density gradients along isobars.  As a consequence of geostrophic
and hydrostatic balance, these gradients would lead to vertical wind
shears \citep{Sun1991}.  These vertical wind shears acting over the
entire atmosphere would likely lead to cloud top winds that would be
highly incompatible with the observed winds.  Thus we considered
methane vertical distributions in which the higher latitudes had
depressed mixing ratios restricted to shallow depths in the upper
troposphere only. As indicated by KT2009, the 2002 spectral
observations did not require that methane depletions extend to great
depths, and \cite{Sro2011occult} showed that shallow depletions were
preferred by the 2002 spectra. We will show here that the 2012
spectral constraints favor relatively shallow methane depletions at
middle latitudes but increasingly deeper depletions at high latitudes.

Here we constructed profiles with shallow \chf depletion using the
``proportionally descended gas'' profiles of \citep{Sro2011occult} in
which the model F1 mixing ratio profile $\alpha (P)$ is dropped down
to increased pressure levels $P'(\alpha)$ using the equation
\begin{eqnarray}
 P' = P\times [1 + (\alpha(P)/\alpha_{d})^{vx}(P_{d}/P_{c}-1)],\label{Eq:deplete}
\end{eqnarray}
for $ P_{tr} <P<P_{d}$,
where $P_{d}$ is the pressure depth at which the revised mixing ratio
$\alpha'(P)=\alpha(P')$ equals the uniform deep mixing ratio
$\alpha_{d}$, $P_{c}$ is the methane condensation pressure before
methane depletion, $P_{tr}$ is the tropopause pressure (100 mb), and
the exponent $vx$ controls the shape of the profile between 100 mb and
$P_{d}$.  
The profiles with $vx=1$ are similar in
form to those adopted by \cite{Kark2011nep}.

\subsection{Compact model fits at 60\deg N vs depletion depth}

To account for the lower methane mixing ratio that is observed at high
latitudes we need to find a methane sink that can create local
depletions.  The most obvious one is methane condensation, which can
cause the low-temperature  above-cloud region to have much lower
methane mixing ratios than the regions below.  When this depleted gas
is mixed downward, it can cause local depressions of the methane
mixing ratio in the upper troposphere.  Thus we expect only the upper
troposphere to be depleted and a significant question is whether we
can constrain the depth of that depression using STIS observations.

\begin{figure}\centering
\includegraphics[width=3.2in]{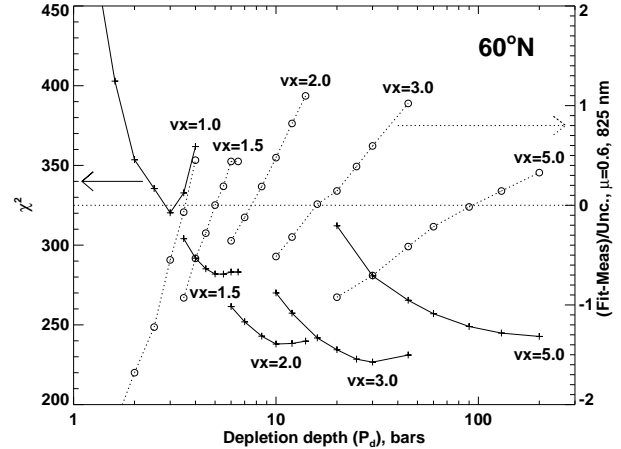}
\caption{Fit quality estimates for compact cloud layer model fits to
  spectra at 60\deg N (\chisq on left axis, 825-nm error on right),
  shown as function of depletion depth $p_d$ for depletion profiles
  with different $vx$ values.  Depletions are relative to an F1 base
  model.}
\label{Fig:60Nvx}
\end{figure}

\begin{figure*}\centering
\includegraphics[width=3.in]{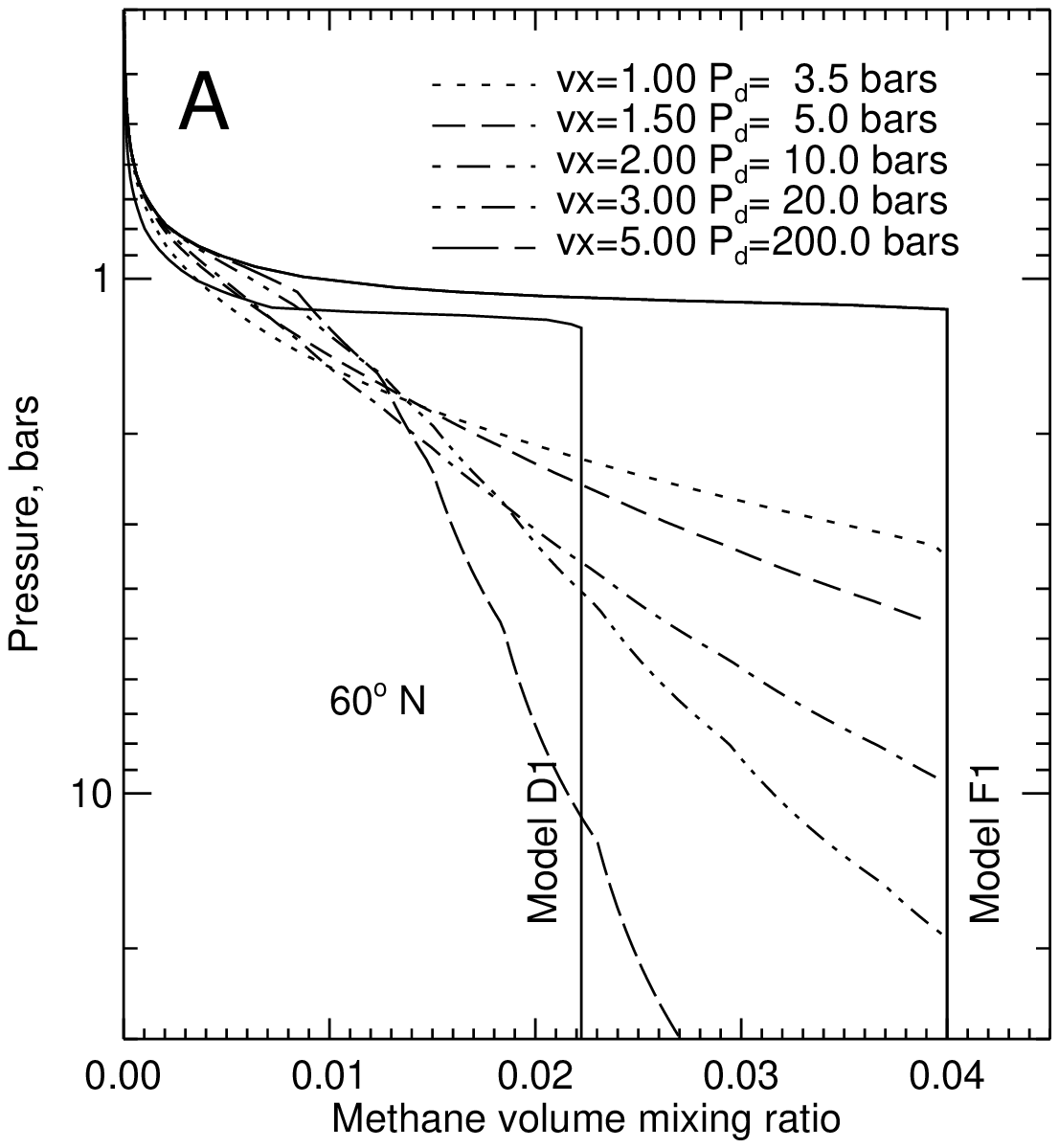}
\includegraphics[width=3.in]{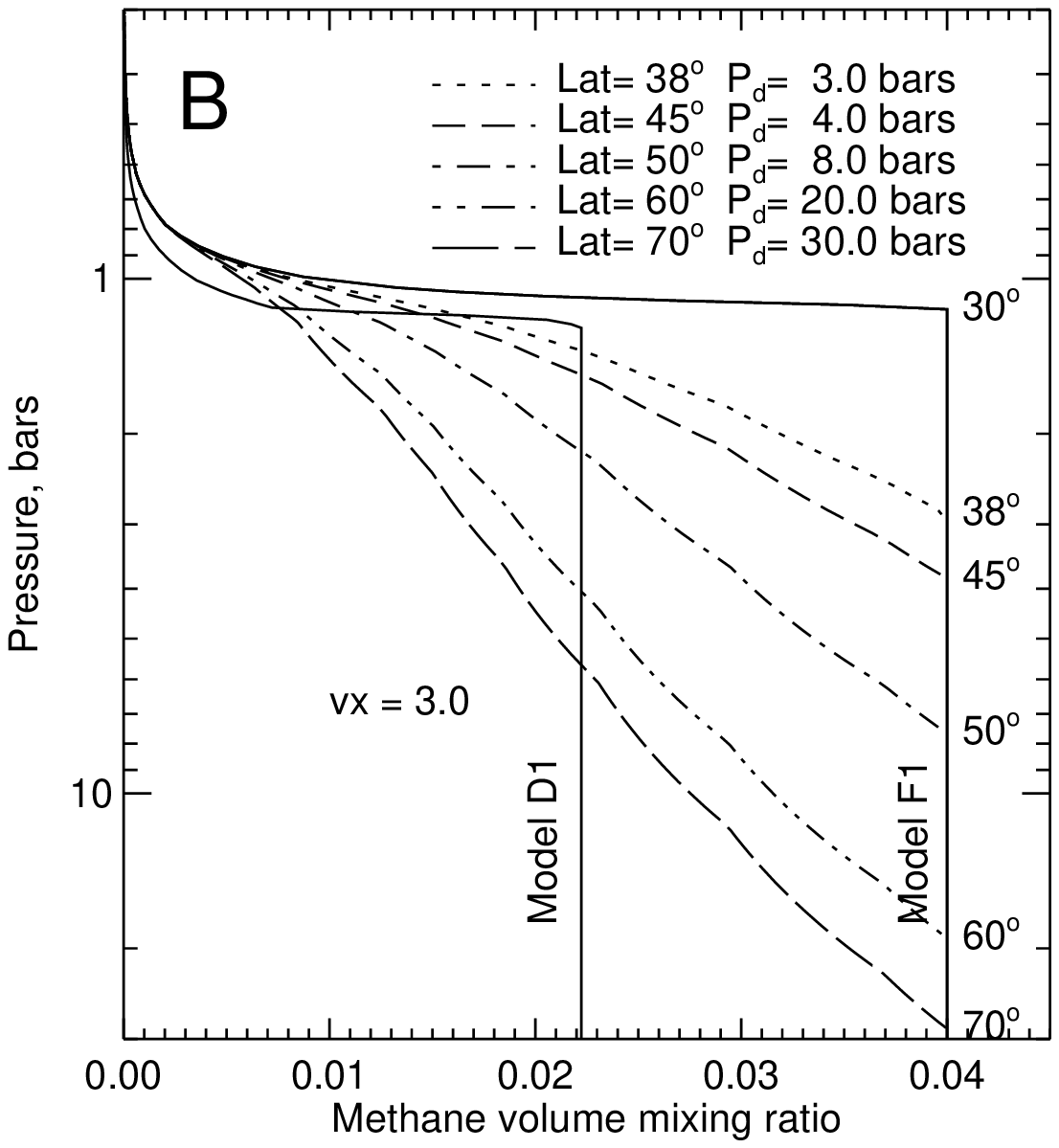}
\caption{Depletion models for 60\deg N with different $vx$ values (A)
and best fit depletion models for different latitudes (B), assuming $vx$=3.0 for
all latitudes.}
\label{Fig:depletions}
\end{figure*}

Here we start with an F1 profile at 60\deg N and then use
Eqn. \ref{Eq:deplete} to deplete the methane above a base pressure
$P_d$ at a rate controlled by a secondary parameter $vx$, which causes
rapid depletion with decreasing pressure when set to small values and
smaller depletion rates when set to larger values.
\cite{Sro2011occult} found that $vx=1$ seemed to be preferred by the
2002 data set, but we found larger values preferred by the 2012 data
set (this might also be true for 2002 if explored more extensively and
with the 2012 calibration function).  We tried a range of $P_d$ values
for $vx$ ranging from 1 to 5.  Subsets of our results are plotted in
Fig.\ \ref{Fig:60Nvx}, which displays \chisq and 825-nm error values
as a function of $P_d$ for different assumed values of $vx$. The
optimum values are summarized in Table\ \ref{Tbl:pdvslat}.  We found
that \chisq minima appear at increasing depths as the depletions
become more gradual with decreasing pressure (for larger values of
$vx$).  The best fit at 60\deg N was found for $vx=3.0$, with $P_d=30$
bars to minimize \chisq and 16 bars to minimize the 825-nm error.
For $vx=2$, the \chisq minimum is near 10 bars and somewhat larger
than the minimum for $vx=3$, while the minimum 825-nm error is found
at 7.5 bars, closer to the \chisq minimum than for the $vx=3$ case.
Clearly, $vx=1$ is a very poor choice. But even that profile fits
better (\chisq = 320) than the D1 profile (\chisq = 360), which is the
best fitting of the occultation consistent profiles with vertically
uniform mixing ratios. The vertical variation of methane VMR for the
best fit $P_d$ value is shown for each $vx$ case in
Fig.\ \ref{Fig:depletions}A.  Note that shallower and deeper
depletions all produce similar mixing ratios near 1.7 bars, and that
greater depletions at depth result in somewhat higher mixing ratios at
pressures below 1.7 bars. It should also be noted that the STIS
measurements aren't sensitive to the large depths implied by the
best-fit $P_d$ values.  It is fitting the shape of the empirical
profile at much lower pressures that constrains the $P_d$ value.  Thus
it would be worthwhile to explore other depletion functions with
different vertical variations.

\subsection{Latitudinal variation in depletion depth and aerosol structure}

To fix the problem of poor spectral fits at latitudes beyond 30\deg N,
we chose $vx=2$ and $vx=3$ vertical variation functions and found the
best-fit value of $P_d$ as a function of latitude for each. At all
latitudes we found that $vx=3$ provided the best fits.  The \chisq and
825-nm error results as a function of $P_d$ for $vx=3$ are plotted in
Fig. 12 of the analysis supplement.  Here we list best-fit $P_d$
values in Table\ \ref{Tbl:pdvslat} and the best-fit aerosol parameters
in Table\ \ref{Tbl:compactF1simvclat}, where the aerosol model is
chosen for $P_d$ values between those giving the smallest \chisq and
those giving the smallest 825-nm error.  The best-fit aerosol results
are also plotted in Fig.\ \ref{Fig:compactvslat} (right panels) for
comparison with the fits using the undepleted F1 profile at all
latitudes. The depletion depth increases from 30\deg N to 70\deg N
(the highest latitude fit) and in this region \chisq and 825-nm errors
have been greatly reduced relative to the prior fits with undepleted
profiles.

One puzzling result is that the cloud layer that we thought was
associated with methane condensation (the UMTC or $m2$ layer in
Fig.\ \ref{Fig:structuremodels}) continues into the polar regions
where methane condensation should not occur because of the decreased
methane mixing ratio.  Perhaps we should not have used the same labels
for $m2$ points at high latitudes that we used at latitudes from
30\deg S to 30\deg N (see right panel of
Fig.\ \ref{Fig:compactvslat}).  In fact, the sharp drop in optical
depth of the $m2$ layer between 20\deg N and 30\deg N might have
continued northward, and what we identified as the $m2$ layer north of
30\degx, might actually be a continuation of the $hg2$ cloud.  In that
case, the layer we identified as the $hg2$ cloud in these northern
fits might actually have a different composition from the
corresponding layer at low latitudes.  Also note that the $hg3$ layer
declines dramatically north of 50\deg N, where $Pd$ exceeds the
assumed depth of the cloud layer.  The polar region possibly has a
completely new set of cloud layers.

\begin{table*}\centering
\caption{Optimum depletion depth based on fit quality and 825-nm error as a function of depletion
profile rate (small $vx$ provides sharper depletions above the depletion depth limit $p_d$).} 
\vspace{0.15in}
\small
\begin{tabular}{|c| c | c c c | c c c|}
\hline
Planetographic &    & \multicolumn{3}{c}{minimum \chisq} &\multicolumn{3}{c}{minimum 825-nm error}\\
Latitude, \deg N & $vx$ & $p_d$ & \chisq & 825-nm error & $p_d$, bars & 825-nm error & \chisq\\
\hline\\[-0.15in]
38 & 2.0 & 2.0 & 215 &-0.50 & 3.0  & 0.28 & 226 \\
38 & 3.0 & 3.0 & 204 &-0.23 & 4.0 & 0.10 & 207 \\
45 & 2.0 & 2.0 & 318 &-0.77 & 3.0  & -.07 & 326 \\
45 & 3.0 & 4.0 & 307 & -0.18 & 4.0 & -0.18 & 307\\
50 & 2.0 & 4   & 239 &-0.01 & 4  &-0.01 & 239 \\
50 & 3.0 & 10  & 225 & 0.57 & 8  & 0.29 & 226 \\   
60 &1.5 &  5.5 & 282 & 0.19 & 5.0  & 0.00 & 282 \\ 
60 &2.0 & 10 & 238 & 0.48 & 7.0  & -0.12 & 252 \\
60 & 3.0 & 30   & 227 & 0.60 & 16   &0.01 & 242 \\ 
60 & 5.0 & 200 & 243 & 0.34 & 90 &  -0.02 &  249 \\
70 & 2.0 & 12  & 253 & 0.50 & 8   & -0.18 & 263 \\ 
70 & 3.0 & 30  & 246 & 0.33 & 20 & -0.09 & 254 \\
\hline 
\end{tabular}\label{Tbl:pdvslat}
\end{table*}

\section{Discussion: Methane as a constraint on meridional motions.} 

\begin{figure}\centering
\includegraphics[width=3.2in]{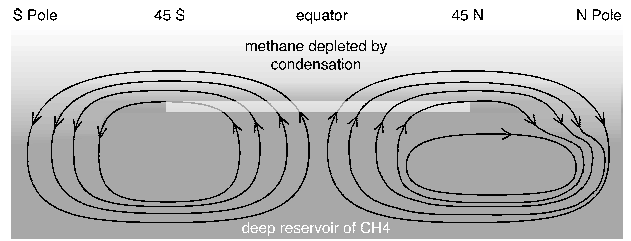}
\caption{Speculative meridional flows in which rising flows at low
  latitudes produce methane condensation (white region) that depletes
  the gas of methane; the depleted gas is transported to high
  latitudes where it descends to reduce the mixing ratio at higher
  pressure. The depletion depth might be limited by the depth of the
  circulation, or by lateral mixing. A very deep cell of this type has
  been inferred from microwave observations (see main text). Note that
  the streamlines of the northern cell provide an example of shallower
  depletions at intermediate high latitudes than those of the southern
  cell, the former option being more consistent with the observed increasing
  methane depletion depth at higher latitudes.}
\label{Fig:circ}
\end{figure}

Fig.\ \ref{Fig:circ} illustrates a mechanism for depleting methane at
high latitudes that is essentially the same as what was previously
suggested by KT2009.
Rising methane-rich gas at
low latitudes is dried out by condensation and sedimentation of
methane ice particles. That dried gas is then transported to high
latitudes, where it begins to descend, bringing down methane depleted
gas, which then gets mixed with methane-rich gas on its return flow.
The figure illustrates two different styles of return flow.
Without lateral eddy mixing across the streamlines, the only restoration
of the methane mixing ratio would be through evaporation of the
precipitating condensed methane at low latitudes.  The depth of the
depletion at high latitudes might be controlled by the depth of
the meridional cell or the depth at which cross-streamline mixing
predominates.
The suggested circulation would promote formation of optically thin
methane clouds (or hazes) at low latitudes, but inhibit methane cloud
formation at high latitudes. This seems consistent with the lack of
observed discrete cloud features south of 45\deg S. However,
we have seen discrete cloud features at high northern latitudes
\citep{Sro2009eqdyn}, which might very well be composed of
something other than methane, given that their cloud tops seem
to be deeper than the methane condensation level \citep{Sro2012polar}.
The general downwelling would also tend to inhibit all condensation
clouds, as sub-condensation mixing ratios would be created by such
motions.  However, this does not rule out localized regions of
upwelling and formation of condensation clouds occupying a small
fractional area.  

The circulation cell in Fig.\ \ref{Fig:circ} would need to be very
deep to be consistent with microwave observations probing the 10-100
bar region of Uranus
\citep{DePater1989Icar82,DePater1991Icar,Hofstadter2007DPS}.  These
reveal a symmetry pattern in which microwave absorbers (NH$_3$,
H$_2$S) are depleted at both high southern and high northern
latitudes, suggesting a non-seasonal equator-to-pole meridional
circulation, with upwelling at low latitudes and down-welling at high
latitudes \citep{DePater2010,DePater1989Icar82,DePater1991Icar,Hofstadter2007DPS}, similar to the circulation
cells illustrated in Fig.\ \ref{Fig:circ}.  If this deep cell extended
to the upper troposphere,
it could be consistent with a shallow CH$_4$ depletion, as long as the
flow at the several bar level was dominated by poleward flow that did
not go through the drying-by-condensation process, as suggested by the
inner streamline in the northern flow pattern in
Fig.\ \ref{Fig:circ}. While our best fits indicate that the largest
fractional depletions occur in the upper troposphere, at higher
latitudes some degree of depletion could extend deeper than the 10-bar
level.  Because these deep meridional cells would likely be dominated
by deep atmospheric conditions, they would probably have the same
symmetry properties as the deep atmosphere (symmetry about the
equator), which would suggest qualitatively that the north and south
polar regions should not look very different. With respect to the
gross character of the methane depletion, they don't look very
different, even though they do exhibit different cloud morphologies.

\begin{figure}\centering
\includegraphics[width=3.2in]{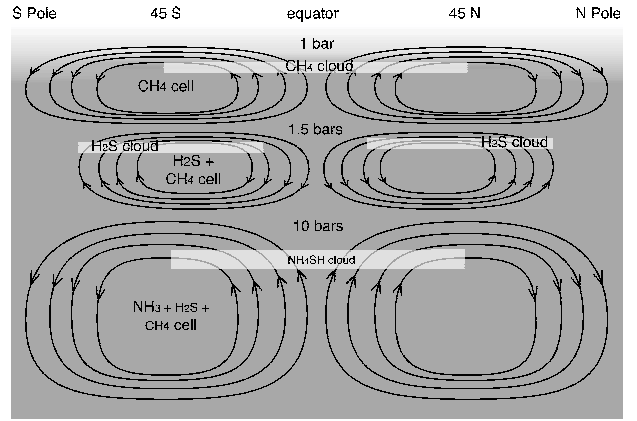}
\caption{Speculated 3-layer cell structure providing equatorward flow
  at the berg level and poleward flow above methane condensation and
  \nhfsh cloud formation levels.  Here cloud formation/condensation
  regions are indicated by light rectangles.}
\label{Fig:circ3}
\end{figure}

A problem with a single deep cell in each hemisphere is that it would
seem to inhibit formation of H$_2$S condensation clouds that are good
candidates for producing the bright bands that form between 38\deg and
58\deg in both hemispheres. In fact, such a simple circulation would
tend to inhibit all condensation clouds at high latitudes, in spite of
the fact that cloud particles of some kind are detected there. A
3-layer circulation cell (Fig.\ \ref{Fig:circ3}) can provide more
opportunities for condensation cloud formation. In this case the deep
cell produces ammonia depletion through formation of an NH$_4$SH cloud
at low latitudes, an intermediate cell produces high latitude
condensation of H$_2$S, and the top cell provides high-latitude
depletion of methane.  This structure is also qualitatively compatible
with the observed equatorward motion of the long-lived and largest
discrete cloud feature seen on Uranus, known as the Berg
\citep{Sro2005dyn,Sro2009eqdyn,DePater2011}.  This feature is thought
to have been at the level of the putative H$_2$S cloud, and in this
model would ride along with the equatorial flow near the 1.5 bar
level.  However, the resemblance between the speculated structure in
Fig.\ \ref{Fig:circ3} and the structure shown in the right panels of
Fig.\ \ref{Fig:compactvslat} is crude at best.
While there seems to
be a change in cloud structure between high and low latitudes, if the
interpretation is restricted to models with condensation clouds only,
it is hard to explain the existence of what seems to be an H$_2$S
cloud at low latitudes where the three-layer cell structure would
inhibit such a cloud.  Furthermore, the drift of the Berg may not be a relevant
constraint on meridional flow. If the cloud features comprising the
Berg were generated by an unseen vortex, its drift may be controlled
more by the vorticity of the zonal flow than by weak meridional flows
\citep{LeBeau1998}.

An additional problem with the above suggested meridional
circulation structures is that they do not appear to be consistent with
Voyager observations. Using thermal emission measurements by the
Voyager IRIS instrument in 1986 \cite{Flasar1987} inferred upwelling
motions at $\pm$30\degx, where relatively lower temperatures were
measured, and subsidence at both equator and poles. The temperatures
were measured in the 60-200 mbar and 600-1000 mbar regions, and the
circulations inferred from a zonally symmetric linear model with
frictional and radiative damping.  But this symmetric pattern is not evident in
the para fraction distribution inferred by \cite{Conrath1998}, which
indicates a seasonal contrast between hemispheres at the time of the
Voyager encounter, with the northern hemisphere having lower
temperatures and higher para fractions, consistent with radiative
cooling and downward motion.  Hadley cell configurations with
hemispherical symmetry are also not consistent with recent numerical
modeling of the seasonal circulation of Uranus by \cite{Sussman2012},
which indicates that there should be tropospheric cross-equatorial
flow peaking near equinox. Another point worth noting is that the
presumed upwelling at low latitudes, which is supported by the
inference of an optically thin methane cloud in those regions, is not
accompanied by an abundance of discrete cloud features at those
latitudes.  Likewise, the observed cloud structure at high northern
latitudes, which appear somewhat deeper than expected for condensed
methane, may be more related to localized upwellings rather than any
broadly defined meridional cells.  This may be similar to the
situation on Jupiter in which a region of overall downwelling (a belt)
contains many examples of localized convection and lightning
\citep{Showman2005}.  It is also conceivable that local increases in
methane humidity might affect aerosol opacity even though the methane
mixing ratio remained below saturation.
The effect of water vapor humidity on the size and scattering
properties of hygroscopic aerosols is significant and well known in
the earth's atmosphere \citep{Pilat1966, Kasten1969}.  It is
conceivable that a haze of hydrocarbon aerosols originally formed in
the stratosphere might have their scattering properties altered in
regions of varying methane humidity, even without the occurrence of
methane condensation.  Such an effect might even give the appearance
of a condensation cloud in regions of local upwelling. However, we do
not know if the background aerosols on Uranus would interact with
methane in the same way as hygroscopic aerosols interact with water
vapor in the earth's atmosphere.

Neptune provides an interesting second example of methane depletion at
high latitudes.  Using 2003 STIS spectra of Neptune,
\cite{Kark2011nep} showed that between 80\deg S and 20\deg N the
variation in Neptune's effective methane mixing ratio was very similar
to that observed for Uranus.  To explain this distribution they also
suggested a mechanism similar to what is illustrated in
Fig.\ \ref{Fig:circ}.  In this case, the 45\deg - 50\deg S latitude
band, one of the most active regions of cloud formation on Neptune
\citep{Kark2011nep500} turns out to be a region where downwelling is
needed to explain the methane depletion results.  However, recent
papers suggest a meridional circulation for Neptune that has upwelling
in this region.  From a re-analysis of Voyager/IRIS 25-50 \mum mapping
of tropospheric temperatures and para-hydrogen disequilibrium
\cite{Fletcher2014nep} suggested a symmetric meridional circulation
with cold air rising at mid-latitudes and warm air sinking at equator
and poles.  Based on a multiwavelength analysis that included near-IR
to microwave observations in 2003, \cite{DePater2014} also detected
warm polar and equatorial regions, where they infer downwelling
motion, and cooler middle latitudes, where they infer upwelling
motion. Such a circulation pattern is inferred to extend to great
depths and would seem to be in conflict with the pattern needed to
produce the observed upper tropospheric methane depletion.  Thus, both
planets seem to have more complicated stories than we are currently
able to explain with simplified models.

\section{Summary and Conclusions}

We observed Uranus with the HST/STIS instrument in 2012, aligning the
instrument's slit parallel to the spin axis of Uranus and stepping the
slit across the face of Uranus from the limb to the center of the planet, building
up half an image with each of 1800 wavelengths from 300.4 to 1020
nm. The main purpose was to constrain the distribution of methane in
the atmosphere of Uranus, taking advantage of the wavelength region
near 825 nm where where hydrogen absorption competes with methane
absorption and displays a clear spectral signature. Our analysis of
STIS observations of Uranus from 2012 and comparisons with similar
2002 observations, as well as analysis of imaging observations from
2007, have led us to the following conclusions.

\begin{enumerate}

\item In 2012 the methane mixing ratio in the upper troposphere of
  Uranus was depleted at high northern latitudes (relative to
  equatorial values), especially beyond 40\deg latitude, to a degree
  very similar to what was inferred for southern high latitudes from
  2002 STIS observations. This is based on (1) direct spectral
  comparisons of STIS latitudinal profiles at wavelengths with similar
  penetration depths but different amounts of hydrogen absorption, (2)
  simplified quantitative modeling of the 815-835 nm spectrum, and (3)
  full radiative transfer modeling.

\item We also found that the north and south depletions were
  simultaneous in 2007, as also suggested by the \cite{Tice2013}
  analysis of 2009 SpeX central meridian spectra, and thus probably
  not a seasonal effect.  Our 2007 result is based on direct spectral
  comparisons near the equinox, using an HST/NICMOS F108N image
  that is sensitive to H$_2$ absorption and a Keck/NIRC2
  PaBeta-filtered image that senses about the same atmospheric level,
  but is dominated by methane absorption.

\item We followed KT2009 in using a simplified model of the 815 nm -
  835 nm spectral region to estimate the relative latitudinal
  variation of the methane volume mixing ratio at 1\deg intervals.
  When this relative variation was absolutely scaled to match
  effective mixing ratios determined by full radiative transfer
  modeling at 10\deg S, the equator, and 60\deg N, we found that the
  effective mixing ratio varied on both large and small spatial
  scales.  At the large scale we found the VMR to increase from
  roughly 2\% within 30-40\deg of the poles to about 4\% within
  20-30\deg of the equator. The 2012 observations suggest an overall
  increase in methane VMR at low latitudes by about 0.5\% relative to
  2002. However, this might be due to a change in the vertical distribution
of aerosols at low latitudes rather than a change in methane VMR.

\item Between 60\deg N and 82\deg N, the simplified model revealed a variation of $\pm$0.005 in
  methane VMR relative to a local mean of 0.020, and a nearly sinusoidal
  variation with a period of about 10\deg in latitude.  A similar
  variation was not seen in south polar regions using the same
  analysis techniques on a similar STIS data set from 2002. Nevertheless,
there is a chance that this feature is not real and needs confirmation
by further measurements.

\item Using Keck2/NIRC2 high-pass filtered H-band images from 16
  August and 4 November 2012, we computed zonally averaged
  brightnesses and compared their variations with latitude with the
  methane VMR variations between 55\deg N and 82\deg N.  We found
  similar patterns in the latitudinal variations of cloud reflectivity
  in August and November images and both had significant negative
  correlations with the methane VMR modulations in the same latitude
  region, suggesting that latitude bands of reduced above-cloud
  methane make the clouds appear brighter due to reduced methane
  absorption.  An alternative interpretation is that the methane
  mixing ratio is relatively constant in this region, with para
  fraction variations (associated with local vertical circulation
  cells) explaining the variation in relative strengths of CH$_4$ and
  H$_2$ absorptions. In this case downwelling correlates with reduced
  cloud reflectivity.

\item At 60\deg N, we tried a variety of vertical variation functions
  for the methane depletion profile, characterized by the exponent
  $vx$, which we varied from 1 to 5.  We found that the depletion
  depth $P_d$ increased as the sharpness of the depletion decreased,
  with best-fit mixing ratios near 1.7 bars being comparable for all
  choices of $vx$.  The best fit of all was obtained for $vx$ = 3,
  with $vx$ = 2 also providing a good fit with somewhat better
  agreement between the $P_d$ that minimized \chisq (10 bars) and the
  $P_d$ that minimized the 825-nm error (7.5 bars).

\item We carried out fits to determine depletion depth as a function
  of latitude, assuming a depletion profile shape defined by $vx=2$
  and also by $vx=3$, with the latter shape providing the best fit at
  all latitudes. We found that $P_d$ increased with latitude,
  beginning just beyond 30\deg N.  At high latitudes the depth of the
  downwelling flow could exceed ten bars or more, although we are only
  sensitive to the upper tropospheric depletions ($P < 10$ bars, as
  indicated by the penetration depth plot of Fig. 2), so that the
  relatively large depletion depths we found may be partly a result of
  the particular empirical function we used in our models. Other
  profile shapes might be able to fit the data without producing as
  great a depth of depletion.

\item Using the depleted profiles to constrain the aerosol parameters,
  we found a lowering in the altitude (increase in base pressure) of
  the $m2$ layer north of 30\deg N (the third aerosol layer from the
  top).  Since the lowered methane mixing ratio also implies that this
  layer can no longer be associated with widespread methane
  condensation (at lower latitudes it is located at the methane
  condensation level), it might here be composed of other materials or
  produced by widely dispersed local convective events, or produced by
  changes in background aerosols due to absorption of methane instead
  of condensation of methane.

\item The association of high-latitude methane depletions with
  descending motions of an equator-to-pole deep Hadley cell does not
  seem to be consistent with the behavior of the detected aerosol
  layers, at least if one ignores other cloud generation mechanisms
  such as sparse local convection.  Both on Uranus and Neptune,
  aerosol layers seem to form in what are thought to be downwelling
  regions on the basis of the effective methane mixing ratio
  determinations.  A three-layer set of circulation cells offers some
  advantages in producing condensation clouds, but also fails to
  provide a good match to the detected aerosol layers.


\item Secular changes between 2002 and 2012 vary with wavelength. At
  continuum wavelengths changes appear to be very small.  When the KT2009
  calibration is adjusted by 3\% to match WFPC2 bandpass filtered
  images, the 2002 I/F value are found to be 2\% greater than
  corresponding 2012 values.  This residual difference may be due
  to calibration uncertainties.  On the other hand, at wavelengths with
  noticeable gas absorption (as in Fig.\ \ref{Fig:latscan3}), the northern hemisphere has
  brightened considerably since 2002, by about 25\% at mid latitudes
  at 827 nm, and the southern hemisphere has darkened, by about 25\%
  at mid latitudes at 827 nm.

\end{enumerate}

\noindent In the future, better constraints on the vertical profile of
methane as a function of latitude could be addressed by additional
modeling work with the 2012 STIS spectra, trying different functional
forms for vertical depletion profiles.  More detailed analysis at more
latitudes using full radiation transfer modeling for both 2002 and
2012 would be useful in clarifying whether low-latitude changes
between 2002 and 2012 are real. Additional STIS observations in
future cycles are also needed to confirm the northern high-latitude
modulations we found in the apparent mixing ratio.  Additional
quantitative constraints might also be derived from analysis of the
vertical wind shears that are implied by the horizontal density
gradients associated with latitudinal compositional gradients.
Additional work with numerical circulation modeling might also be
productive in understanding how the methane mixing ratio affects and
is affected by atmospheric circulation patterns.

\section*{Acknowledgments}\addcontentsline{toc}{section}{Acknowledgments}

This research was supported primarily by grants from the Space
Telescope Science Institute, managed by AURA. GO-12894.01-A supported
LAS and PMF. Partial support was provided by NASA Planetary Astronomy
Grant NNX13AH65G (LAS and PMF). EK, HBH, IdP, and KAR also acknowledge
support from STScI grants under GO-12894.  We thank staff at the
W. M. Keck Observatory, which is made possible by the generous
financial support of the W. M. Keck Foundation.  We thank those of
Hawaiian ancestry on whose sacred mountain we are privileged to be
guests. Without their generous hospitality none of our groundbased
observations would have been possible.  We also thank
Robert West and an anonymous reviewer who provided very thorough
reviews and constructive suggestions for improving the paper.

\addcontentsline{toc}{section}{References}


\begin{thebibliography}{33}
\expandafter\ifx\csname natexlab\endcsname\relax\def\natexlab#1{#1}\fi
\expandafter\ifx\csname url\endcsname\relax
  \def\url#1{\texttt{#1}}\fi
\expandafter\ifx\csname urlprefix\endcsname\relax\def\urlprefix{URL }\fi

\bibitem[{{Acton}(1996)}]{Acton1996}
{Acton}, C.~H., 1996. {Ancillary data services of NASA's Navigation and
  Ancillary Information Facility}. Planet. and Space Sci. 44, 65--70.

\bibitem[{{Borysow} et~al.(2000){Borysow}, {Borysow}, and {Fu}}]{Borysow2000}
{Borysow}, A., {Borysow}, J., {Fu}, Y., 2000. {Semi-empirical model of
  collision-induced absorption spectra of H$_2$-H$_2$ complexes in the second
  overtone band of hydrogen at temperatures from 50 to 500 K}. Icarus 145,
  601--608.

\bibitem[{{Colina} et~al.(1996){Colina}, {Bohlin}, and {Castelli}}]{Colina1996}
{Colina}, L., {Bohlin}, R.~C., {Castelli}, F., 1996. {The 0.12-2.5 micron
  Absolute Flux Distribution of the Sun for Comparison With Solar Analog
  Stars}. \aj 112, 307--315.

\bibitem[{{Conrath} et~al.(1998){Conrath}, {Gierasch}, and
  {Ustinov}}]{Conrath1998}
{Conrath}, B.~J., {Gierasch}, P.~J., {Ustinov}, E.~A., 1998. {Thermal Structure
  and Para Hydrogen Fraction on the Outer Planets from Voyager IRIS
  Measurements}. Icarus 135, 501--517.

\bibitem[{{de Pater} et~al.(2014){de Pater}, {Fletcher}, {Luszcz-Cook},
  {DeBoer}, {Butler}, {Hammel}, {Sitko}, {Orton}, and {Marcus}}]{DePater2014}
{de Pater}, I., {Fletcher}, L.~N., {Luszcz-Cook}, S., {DeBoer}, D., {Butler},
  B., {Hammel}, H.~B., {Sitko}, M.~L., {Orton}, G.~O., {Marcus}, P.~S., 2014.
  {Neptune's global circulation deduced from multi-wavelength observations}.
  Icarus 000, submitted.

\bibitem[{{de Pater} and {Lissauer}(2010)}]{DePater2010}
{de Pater}, I., {Lissauer}, J.~L., 2010. {Planetary Sciences, 2nd ed.}
  Cambridge University Press.

\bibitem[{{de Pater} et~al.(1989){de Pater}, {Romani}, and
  {Atreya}}]{DePater1989Icar82}
{de Pater}, I., {Romani}, P.~N., {Atreya}, S.~K., 1989. {Uranus deep atmosphere
  revealed}. Icarus 82, 288--313.

\bibitem[{{de Pater} et~al.(1991){de Pater}, {Romani}, and
  {Atreya}}]{DePater1991Icar}
{de Pater}, I., {Romani}, P.~N., {Atreya}, S.~K., 1991. {Possible microwave
  absorption by H2S gas in Uranus' and Neptune's atmospheres}. Icarus 91,
  220--233.

\bibitem[{{de Pater} et~al.(2011){de Pater}, {Sromovsky}, {Hammel}, {Fry},
  {LeBeau}, {Rages}, {Showalter}, and {Matthews}}]{DePater2011}
{de Pater}, I., {Sromovsky}, L., {Hammel}, H.~B., {Fry}, P.~M., {LeBeau},
  R.~P., {Rages}, K.~A., {Showalter}, M.~R., {Matthews}, K., 2011.
  {Post-equinox Observations of Uranus: Berg's Evolution, vertical structure,
  and track towards the qquator}. Icarus 215, 332--345.

\bibitem[{{Flasar} et~al.(1987){Flasar}, {Conrath}, {Pirraglia}, and
  {Gierasch}}]{Flasar1987}
{Flasar}, F.~M., {Conrath}, B.~J., {Pirraglia}, J.~A., {Gierasch}, P.~J., 1987.
  {Voyager infrared observations of Uranus' atmosphere - Thermal structure and
  dynamics}. \jgr 92, 15011--15018.

\bibitem[{{Fletcher} et~al.(2014){Fletcher}, {de Pater}, {Orton}, {Hammel},
  {Sitko}, and {Irwin}}]{Fletcher2014nep}
{Fletcher}, L.~N., {de Pater}, I., {Orton}, G.~S., {Hammel}, H.~B., {Sitko},
  M.~L., {Irwin}, P.~G.~J., 2014. {Neptune at summer solstice: Zonal mean
  temperatures from ground-based observations, 2003-2007}. Icarus 231,
  146--167.

\bibitem[{{Hernandez} et~al.(2012){Hernandez}, {Aloisi}, {Bohlin}, {Bostroem},
  {Diaz}, {Dixon}, {Ely}, {Goudfrooij}, {Hodge}, {Lennon}, {Long}, {Niemi},
  {Osten}, {Proffitt}, {Walborn}, {Wheeler}, {York}, and
  {Zheng}}]{Hernandez2012}
{Hernandez}, S., {Aloisi}, A., {Bohlin}, R., {Bostroem}, A., {Diaz}, R.,
  {Dixon}, V., {Ely}, J., {Goudfrooij}, P., {Hodge}, P., {Lennon}, D., {Long},
  C., {Niemi}, S., {Osten}, R., {Proffitt}, C., {Walborn}, N., {Wheeler}, T.,
  {York}, B., {Zheng}, W., 2012. STIS Instrument Handbook, Version 12.0,
  (Baltimore: STScI). Space Telescope Science Institute, Baltimore, Maryland.

\bibitem[{{Hofstadter} et~al.(2007){Hofstadter}, {Butler}, and
  {Gurwell}}]{Hofstadter2007DPS}
{Hofstadter}, M.~D., {Butler}, B.~J., {Gurwell}, M.~A., 2007. {Imaging Uranus
  at Submillimeter to Centimeter Wavelengths}. Bull. Am. Astron. Soc. 39, 424.

\bibitem[{{Karkoschka}(2011)}]{Kark2011nep500}
{Karkoschka}, E., 2011. {Neptune's cloud and haze variations 1994-2008 from 500
  HST-WFPC2 images}. Icarus 215, 759--773.

\bibitem[{{Karkoschka} and {Tomasko}(2009)}]{Kark2009IcarusSTIS}
{Karkoschka}, E., {Tomasko}, M., 2009. {The haze and methane distributions on
  Uranus from HST-STIS spectroscopy}. Icarus 202, 287--309.

\bibitem[{{Karkoschka} and {Tomasko}(2011)}]{Kark2011nep}
{Karkoschka}, E., {Tomasko}, M.~G., 2011. {The haze and methane distributions
  on Neptune from HST-STIS spectroscopy}. Icarus 211, 780--797.

\bibitem[{{Kasten}(1969)}]{Kasten1969}
{Kasten}, F., 1969. {Visibility forecast in the phase of pre-condensation}.
  Tellus, 631--635.

\bibitem[{{Krist}(1995)}]{Krist1995}
{Krist}, J., 1995. {Simulation of HST PSFs using Tiny Tim}. In: {Shaw}, R.~A.,
  {Payne}, H.~E., {Hayes}, J.~J.~E. (Eds.), Astronomical Data Analysis Software
  and Systems IV. Vol.~77 of Astronomical Society of the Pacific Conference
  Series. pp. 349--352.

\bibitem[{{Lebeau} and {Dowling}(1998)}]{LeBeau1998}
{Lebeau}, R.~P., {Dowling}, T.~E., 1998. {EPIC Simulations of Time-Dependent,
  Three-Dimensional Vortices with Application to Neptune's Great Dark Spot}.
  Icarus 132, 239--265.

\bibitem[{{Lindal} et~al.(1987){Lindal}, {Lyons}, {Sweetnam}, {Eshleman}, and
  {Hinson}}]{Lindal1987}
{Lindal}, G.~F., {Lyons}, J.~R., {Sweetnam}, D.~N., {Eshleman}, V.~R.,
  {Hinson}, D.~P., 1987. {The atmosphere of Uranus - Results of radio
  occultation measurements with Voyager 2}. J. Geophys. Res. 92, 14987--15001.

\bibitem[{{Pilat} and {Charlson}(1966)}]{Pilat1966}
{Pilat}, M.~J., {Charlson}, R.~J., 1966. {Theoretical and optical studies of
  humdity effects on the size distribution of a hygroscopic aerosol}. Journal
  de Rcherches Atmosph\`eriques, 165--170.

\bibitem[{{Press} et~al.(1992){Press}, {Teukolsky}, {Vetterling}, and
  {Flannery}}]{Press1992}
{Press}, W.~H., {Teukolsky}, S.~A., {Vetterling}, W.~T., {Flannery}, B.~P.,
  1992. {Numerical recipes in FORTRAN. The art of scientific computing, 2nd
  ed.} Cambridge: University Press.

\bibitem[{{Showman} and {de Pater}(2005)}]{Showman2005}
{Showman}, A.~P., {de Pater}, I., 2005. {Dynamical implications of Jupiter's
  tropospheric ammonia abundance}. Icarus 174, 192--204.

\bibitem[{{Sromovsky}(2005{\natexlab{a}})}]{Sro2005raman}
{Sromovsky}, L.~A., 2005{\natexlab{a}}. {Accurate and approximate calculations
  of Raman scattering in the atmosphere of Neptune}. Icarus 173, 254--283.

\bibitem[{{Sromovsky}(2005{\natexlab{b}})}]{Sro2005pol}
{Sromovsky}, L.~A., 2005{\natexlab{b}}. {Effects of Rayleigh-scattering
  polarization on reflected intensity: a fast and accurate approximation method
  for atmospheres with aerosols}. Icarus 173, 284--294.

\bibitem[{{Sromovsky} and {Fry}(2005)}]{Sro2005dyn}
{Sromovsky}, L.~A., {Fry}, P.~M., 2005. {Dynamics of cloud features on Uranus}.
  Icarus 179, 459--484.

\bibitem[{{Sromovsky} and {Fry}(2010)}]{Sro2010iso}
{Sromovsky}, L.~A., {Fry}, P.~M., 2010. {The source of 3-{$\mu$}m absorption in
  Jupiter's clouds: Reanalysis of ISO observations using new NH$_{3}$
  absorption models}. Icarus 210, 211--229.

\bibitem[{{Sromovsky} et~al.(2009){Sromovsky}, {Fry}, {Hammel}, {Ahue}, {de
  Pater}, {Rages}, {Showalter}, and {van Dam}}]{Sro2009eqdyn}
{Sromovsky}, L.~A., {Fry}, P.~M., {Hammel}, H.~B., {Ahue}, W.~M., {de Pater},
  I., {Rages}, K.~A., {Showalter}, M.~R., {van Dam}, M.~A., 2009. {Uranus at
  equinox: Cloud morphology and dynamics}. Icarus 203, 265--286.

\bibitem[{{Sromovsky} et~al.(2012){Sromovsky}, {Fry}, {Hammel}, {de Pater}, and
  {Rages}}]{Sro2012polar}
{Sromovsky}, L.~A., {Fry}, P.~M., {Hammel}, H.~B., {de Pater}, I., {Rages},
  K.~A., 2012. {Post-equinox dynamics and polar cloud structure on Uranus}.
  Icarus 220, 694--712.

\bibitem[{{Sromovsky} et~al.(2011){Sromovsky}, {Fry}, and
  {Kim}}]{Sro2011occult}
{Sromovsky}, L.~A., {Fry}, P.~M., {Kim}, J.~H., 2011. {Methane on Uranus: The
  case for a compact CH$_4$ cloud layer at low latitudes and a severe CH$_4$
  depletion at high latitudes based on re-analysis of Voyager occultation
  measurements and STIS spectroscopy.} Icarus 215, 292--312.

\bibitem[{{Sun} et~al.(1991){Sun}, {Schubert}, and {Stoker}}]{Sun1991}
{Sun}, Z., {Schubert}, G., {Stoker}, C.~R., 1991. {Thermal and humidity winds
  in outer planet atmospheres}. Icarus 91, 154--160.

\bibitem[{{Sussman} et~al.(2012){Sussman}, {Dowling}, {Greathouse}, and
  {Chanover}}]{Sussman2012}
{Sussman}, M., {Dowling}, T.~E., {Greathouse}, T.~K., {Chanover}, N.~J., 2012.
  {Seasonal Circulation Modeling of Uranus}. In: AAS/Division for Planetary
  Sciences Meeting Abstracts. Vol.~44 of AAS/Division for Planetary Sciences
  Meeting Abstracts. p. \#504.02.

\bibitem[{{Tice} et~al.(2013){Tice}, {Irwin}, {Fletcher}, {Teanby}, {Hurley},
  {Orton}, and {Davis}}]{Tice2013}
{Tice}, D.~S., {Irwin}, P.~G.~J., {Fletcher}, L.~N., {Teanby}, N.~A., {Hurley},
  J., {Orton}, G.~S., {Davis}, G.~R., 2013. {Uranus' cloud particle properties
  and latitudinal methane variation from IRTF SpeX observations}. Icarus 223,
  684--698.

\end{thebibliography}


\section*{Supplemental material.}

 An expanded discussion of analysis techniques and results is provided
 in the file ura\_stis\_analysis\_supplement.pdf, which can be found
 on line at
 http://www.ssec.wisc.edu/planetary/uranus/onlinedata/ura2012stis/.
The hyperspectral
 cube containing calibrated I/F values as a function of wavelength and
 location, with a navigation back plane that provides viewing geometry
 and latitude-longitude coordinates for each pixel, is also provided there in
 ura\_stis2012\_datacube.fits.
A detailed explanation of the
contents of this file can be found in the file
README\_SUPPLEMENTAL.TXT.  A sample IDL program that reads the cube
file, plots a monochromatic image, extracts data from a particular
location on the disc, and plots a spectrum, is provided in the file
stis\_cube\_read\_example.pro.  The IDL astronomy library will be needed to
read the data cube, which is in the FITS format.

\end{document}